\documentclass[aps,pra,twocolumn,groupedaddress,floatfix,superscriptaddress]{revtex4-1}

\usepackage{amsmath}
\usepackage{amssymb}
\usepackage{graphicx}
\usepackage{float}
\usepackage[english]{babel}
\usepackage{dsfont}
\usepackage{epstopdf}
\usepackage{soul}
\usepackage{color}
\usepackage{cancel}
\usepackage{braket}
\usepackage{subcaption}
\usepackage{tikz}
\usepackage{lipsum} 
\usepackage[colorlinks=true,
            linkcolor=magenta,
            urlcolor=blue,
            citecolor=blue]{hyperref}
\usepackage[normalem]{ulem}
\usepackage{ragged2e}
\usepackage{braket}
\usepackage[dvipsnames]{xcolor}
\usepackage{comment}

\def\be{\begin{equation}}
\def\ee{\end{equation}} 
\def\bea{\begin{eqnarray}}
\def\eea{\end{eqnarray}} 
\def\ba{\begin{array}} 
\def\ea{\end{array}}

\def\rm{\mathrm}

\newcommand{\edo}[1] {\textcolor{ForestGreen}{#1}}

\emergencystretch=\maxdimen
\begin{document}

\title{Trimer Dynamics in Floquet-driven arrays of Rydberg Atoms}

\author{Edoardo Tiburzi}
\affiliation{Dipartimento di Fisica e Astronomia ``G. Galilei", Università degli Studi di Padova, I-35131, Padova, Italy}
\affiliation{INFN Istituto Nazionale di Fisica Nucleare, Sezione di Padova, I-35131, Padova, Italy}
\author{Lorenzo Maffi}
\affiliation{Dipartimento di Fisica e Astronomia ``G. Galilei", Università degli Studi di Padova, I-35131, Padova, Italy}
\affiliation{INFN Istituto Nazionale di Fisica Nucleare, Sezione di Padova, I-35131, Padova, Italy}
\author{Luca Dell'Anna}
\affiliation{Dipartimento di Fisica e Astronomia ``G. Galilei", Università degli Studi di Padova, I-35131, Padova, Italy}
\affiliation{INFN Istituto Nazionale di Fisica Nucleare, Sezione di Padova, I-35131, Padova, Italy}
\affiliation{Padua Quantum Technologies Research Center (QTECH), Padova, Italy}
\author{Marco Di Liberto}
\affiliation{Dipartimento di Fisica e Astronomia ``G. Galilei", Università degli Studi di Padova, I-35131, Padova, Italy}
\affiliation{INFN Istituto Nazionale di Fisica Nucleare, Sezione di Padova, I-35131, Padova, Italy}
\affiliation{Padua Quantum Technologies Research Center (QTECH), Padova, Italy}

\begin{abstract}
We analyze the WAHUHA Floquet protocol recently applied to arrays of Rydberg atoms and derive beyond-leading-order corrections in the high-frequency expansion of the effective spin theory. 
We find that an appropriate choice of the pulses times can enforce an approximate symmetry corresponding to the conservation of the total magnetization.  
The interaction channels emerging from higher-order Floquet terms affect three-body bound states (\emph{trimers}), which gain a significant mobility. 
We estimate the corresponding enhancement in 1D spin chains and conclude that their dynamics is within experimental reach. 
Detrimental effects due to the proliferation of particles outside of the trimer magnetization sector are found to occur and spread on time-scales slower than the trimer propagation.
We further show that long-range interactions enhance trimer propagation and that two-dimensional triangular geometries can host energetically isolated trimer bands, providing a possible route to reduce resonant mixing with higher-magnetization sectors.
Our results establish a concrete route to realizing mobile multiparticle bound states in Floquet-engineered Rydberg platforms.
\end{abstract}

\maketitle

\section{Introduction}

Spin models provide a paradigmatic framework to describe many-body systems \cite{sachdev2023}.
Recent developments in quantum simulation platforms have opened avenues for their programmable realization with tunable interactions and geometries \cite{Georgescu2014}. 
In this context, spin models have been employed to capture effective quantum properties of systems described by Hubbard models, for example, with ultracold mobile atoms \cite{Duan2003, Gross2013,Greif2013, Gross2016, Greiner2017}.  
Furthermore, they emerge as main description for arrays of long-range interacting atoms, as dipolar atoms and polar molecules \cite{Trefzger2011,Baranov2012}, or Rydberg atoms \cite{saffman_quantum_2010, browaeys_many-body_2020}.
In particular, strong Rydberg-Rydberg interactions have enabled the experimental realization of paradigmatic spin-1/2 Hamiltonians, such as the Ising, XY, and XXZ models, in one and two dimensions by encoding the spin degree of freedom in a pair of atomic levels \cite{lienhard_realization_2020}. 

These long-range interacting systems have made possible to explore the ground state properties of spin models, including quantum phase transitions \cite{Keesling2019, Lukin2021}, topological states \cite{de_leseleuc_observation_2019,Semeghini2021} and low-energy excitations, for example, single magnons, collective excitations or domain walls \cite{Bernien2017, Manovitz2025}. 
In this context, bound states can also appear in the excitation spectrum \cite{Gross2013, ganahl_2012, Petrosyan2018, macri_bound_2021, kranzl_2023, kim_2024, kuang2026dynamics} at the cost of a slower time evolution due to their higher mass. 
Two-particle bound states have been largely investigated with synthetic matter systems with onsite interactions, as for the case of Hubbard models \cite{Winkler2006, tai_microscopy_2017}. 
Systems with long-range interactions however display distinct features. 
They can manifest a critical interaction strength for the existence of stable bound states \cite{Valiente2009}, recently observed with Rydberg dressed bosonic atoms \cite{weckesser_realization_2024}. 
The interplay of long-range strong interactions and lattice geometry can lead to bound states with emergent topological or localization properties absent in the noninteracting regime \cite{Salerno2020,Santos2021}.
Recently, more complex bound states or excitations, like string excitations or domain walls \cite{misguich_2017, Surace2020, Pavesic2025, Cuadra2025}
have been the focus of investigation due to their exotic dynamical properties, including the connection to gauge theories and confinement \cite{calabrese2017}.

While native models of the atomic platform already provide a rich playground to study spin models and their few particle or collective excitations, it is crucial to design schemes that allow a higher degree of control of these many-body systems. 
Floquet engineering has emerged as a versatile method to manipulate atomic systems and their quantum Hamiltonians via time-dependent periodic drives \cite{goldman_periodically_2014, Eckardt2017}. 
In the case of mobile ground state atoms, Floquet techniques have been employed to control many-body transport properties across phase transitions \cite{Eckardt2005,Zenesini2009}, engineer gauge fields \cite{Aidelsburger2011, Struck2012, Goldman2014}, induce topological band structures \cite{abanin_2015, Goldman2016,Cooper2019} and design interacting models \cite{Rapp2012,DiLiberto2014}.
In the context of Rydberg atoms, recent progress has shown that Floquet methods can conveniently be employed to manipulate interactions beyond the limitations of the native models \cite{Zhao2023, Koyluoglu2025}, for example to engineer the Kitaev honeycomb model and its non-Abelian topological properties \cite{Kalinowski2023, Sun2023, Evered2025}.
A particularly convenient driving scheme, known as WAHUHA pulse sequence \cite{waugh_approach_1968}, has been employed to engineer the XYZ Hamiltonian \cite{geier_floquet_2021, scholl_microwave-engineering_2022} at leading order in the Floquet high-frequency expansion.
Tunability of time pulses allows to tune the model in a parameter region where magnetization is conserved. 
In this regime, the model exhibits an $SO(2)$ symmetry and is described by an XXZ spin model that can be tuned to the rotational invariant Heisenberg limit.
However, beyond-leading-order corrections are expected, in general, to generate a proliferation of spin-spin interaction terms that can break this symmetry.

In this work, we investigate  the WAHUHA protocol for long-range interacting Rydberg atoms described by the XY model. 
We derive next-to-leading order Floquet corrections up to order $1/\omega^2$, with $\omega$ the driving frequency, and we find a regime of pulse times that removes all terms violating the conservation of magnetization except for one. 
We then study their impact on the dynamics of three-body bound states (\emph{trimers}) \cite{valiente_2010, Santos2021, giudice_2022}. 
Three-particle bound states are expected to have a very large mass in the XXZ model, which hinders the possibility of observing them in experiments.
Here we show that WAHUHA Floquet engineering provides additional  channels that significantly increase the mobility of trimers.
We also investigate the role of long-range couplings showing that they contribute positively to the trimer mobility. Moreover two-dimensional lattice geometries can help mitigating the proliferation of states from other unwanted magnetization sectors.
Our findings highlight the potential of Floquet-engineered Rydberg systems for studying exotic few-body excitations and correlated dynamics in strongly-interacting quantum matter.

The paper is organized as follows. 
In Sec.~\ref{sec:WAHUHA} we introduce the WAHUHA Floquet driving protocol and derive the corresponding effective Hamiltonian up to second order in the high-frequency expansion. We then specialize to a one-dimensional chain with nearest-neighbor couplings and choose the pulse durations such as to enforce an approximate $SO(2)$  symmetry. In this regime, the effective Hamiltonian acquires a particularly transparent form, in which different contributions can be directly identified as nearest- and next-nearest-neighbor hopping, correlated hopping and multispin processes, making their impact on the mobility of magnons, dimers and trimers explicit. 
In Sec.~\ref{sec:trimer_dynamics} we discuss the formation and dynamics of three-body bound states (trimers) in one-dimensional chains, and analyze how the Floquet-induced terms modify their dispersion and propagation compared to magnons and dimers. 
In Sec.~\ref{sec:LongRange2D} we extend our analysis to systems with long-range interactions and to two-dimensional lattice geometries, showing how these ingredients enhance trimer mobility and at the same time suppress unwanted mixing between different magnetization sectors. 
Finally, in Sec.~\ref{sec:conclusions} we summarize our results and outline perspectives for experimental implementations and future developments. 
Technical details on the Floquet expansion and on the structure of the effective couplings are collected in the Appendices.

\section{WAHUHA Floquet driving}
\label{sec:WAHUHA}
\subsection{Effective theory: Leading order}

\begin{figure}[!t]
    \centering
    \includegraphics[width=\linewidth]{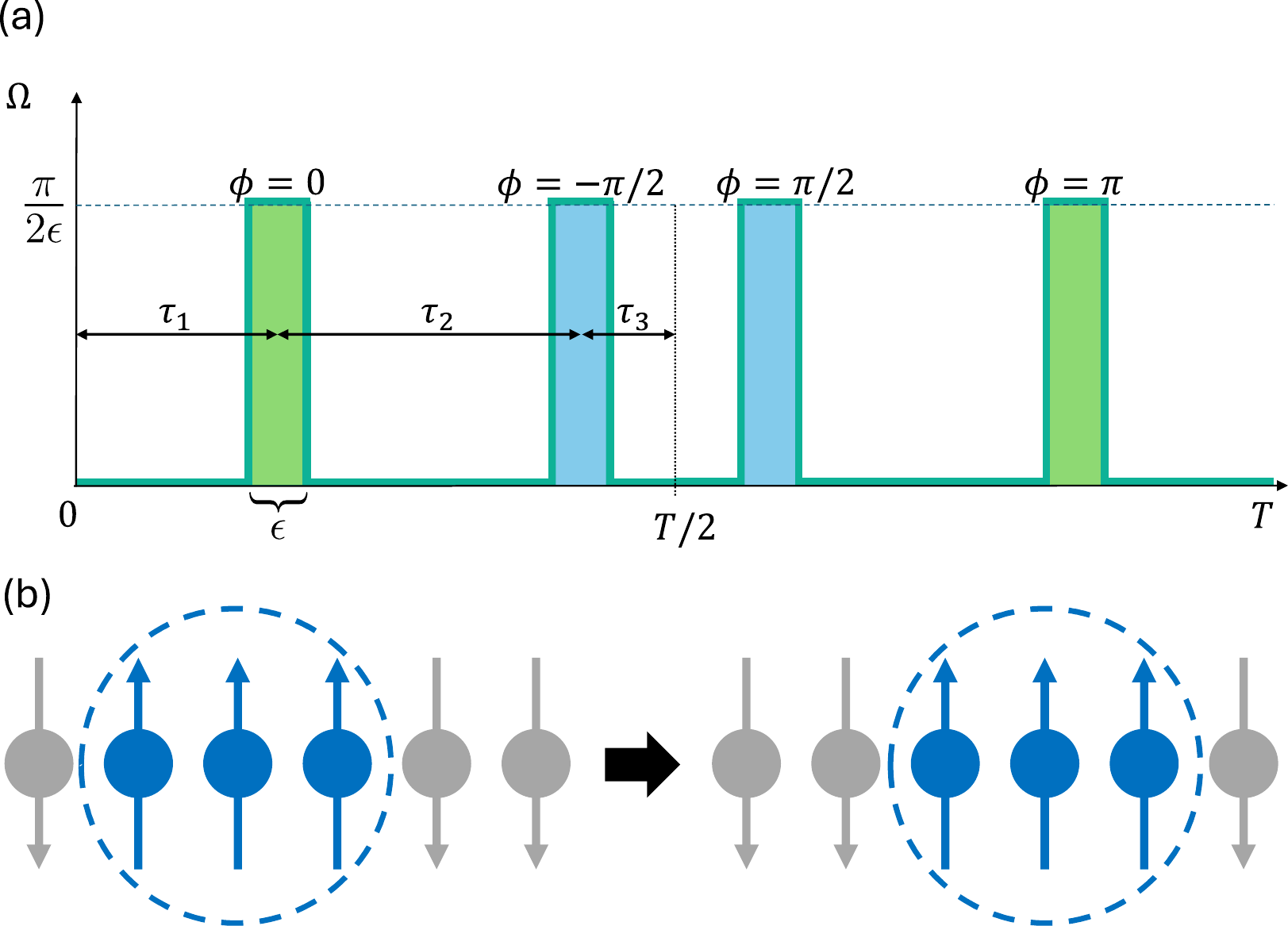}
    \caption{
    \justifying Floquet pulse scheme and trimer dynamics.
    \textbf{(a)} Schematic representation of the WAHUHA Floquet driving sequence, consisting of four $\pi/2$ pulses applied at times $t_1, t_2, t_3, t_4$ with phases $\phi = 0, -\pi/2, \pi/2, \pi$, separated by intervals $\tau_1, \tau_2, 2\tau_3$, such that $\tau_1 + \tau_2 + \tau_3 = T/2$. For square pulses of duration $\epsilon$, the pulse amplitude is $\Omega=\pi/(2\epsilon)$, so that the pulse area is $\pi/2$. This protocol effectively engineers anisotropic spin interactions by averaging out undesired terms. \textbf{(b)} Illustration of a trimer (three adjacent spin excitations forming a bound state) propagating coherently along a one-dimensional spin chain.}
    \label{fig:wahuha+trimer}
\end{figure}

We study the Floquet dynamics of a periodically driven spin system described by the Hamiltonian:
\begin{equation}
\label{eq: full-ham}
    H_{\text{driven}} = H_{\text{XY}} + W(t)\,,
\end{equation}
where 
\begin{equation}
    \label{eq:XY}
    H_{\text{XY}}=\frac{1}{2} \sum_{i \neq j} J_{i j}\left(\sigma_i^x \sigma_j^x+\sigma_i^y \sigma_j^y\right)\,,
\end{equation}
is the native XY Hamiltonian of the spin system at hand, for example arising from dipolar couplings between Rydberg atoms \cite{browaeys_many-body_2020}. The periodic drive $W(t)$ follows the WAHUHA protocol \cite{waugh_approach_1968, geier_floquet_2021, scholl_microwave-engineering_2022}, 
notably realized by the sequence of four $\pi/2$ pulses with fixed phases $\phi = \{0, -\pi/2, \pi/2, \pi\}$
and separated by time intervals $ \tau_1, \tau_2, 2\tau_3 $ such that $\tau_1+ \tau_2 + \tau_3 = T/2$, with $T$ being the driving period (see Fig.~\ref{fig:wahuha+trimer}(a)).
A $\pi/2$ pulse is defined by its pulse area, $\int_{\mathrm{pulse}} dt\,\Omega(t)=\pi/2$. For square pulses of duration $\epsilon$, this corresponds to an amplitude $\Omega=\pi/(2\epsilon)$, as schematically indicated in Fig.~\ref{fig:wahuha+trimer}(a). In the following, we work in the ideal-pulse limit $\epsilon \ll T$, where the pulse duration is neglected with respect to the free-evolution intervals.

In the rotating frame defined by the explicit form of the drive
\begin{equation}
    W(t) = \frac{\hbar \Omega(t)}{2} \sum_i \left[\cos \phi(t) \sigma_i^x+\sin \phi(t) \sigma_i^y\,\right] ,
    \label{eq:pulse}
\end{equation}
the Hamiltonian becomes:
\begin{align}
    \nonumber H'(t) &= U_{\text{drive}}^\dagger(t) H_{\text{driven}} U_{\text{drive}}(t) - i\hbar U_{\text{drive}}^\dagger(t)\frac{dU_{\text{drive}}(t)}{dt}\\
    &= U_{\text{drive}}^\dagger(t) H_{\text{XY}} U_{\text{drive}}(t) ,
    \label{eq:rotated_ham}
\end{align}
where $ U_{\text{drive}}(t)=\mathcal{T}e^{-\frac{i}{\hbar}\int_{0}^{t}d\tau W(\tau)} $ is the time-evolution operator.
According to standard Floquet theory \cite{goldman_periodically_2014, Eckardt2017} (see Appendix \ref{appendix_floquet}) we can describe the effective dynamical behaviour of the system by a systematic high-frequency expansion in $\omega\equiv 2\pi/T$ leading to a time-independent effective Hamiltonian $H_\text{eff}$. 
At leading order, we obtain
\begin{equation}
\label{eq: flq-exp}
    H_\text{eff} = \sum_n \frac{1}{\omega^n} H_\text{eff}^{(n)} \approx H_0 + \mathcal{O}(1/\omega^2),
\end{equation}
where $ H_0 $ was studied in Refs.~\cite{geier_floquet_2021,scholl_microwave-engineering_2022} and it is given by:
\begin{equation}
    \begin{split}
        H_0 =  \sum_{i \ne j} \frac{J_{ij}}{T} &\left[\right. (\tau_1 + \tau_2) \sigma_i^x \sigma_j^x + (\tau_1 + \tau_3) \sigma_i^y \sigma_j^y   \\
        &\left. +(\tau_2 + \tau_3) \sigma_i^z \sigma_j^z \right].
    \end{split}
\label{eq:h0}
\end{equation}
The results above show that this Floquet protocol provides a flexible scheme for engineering a generic XYZ spin model.
By tuning the times $\tau_1$, $\tau_2$ and $\tau_3$, it is possible to obtain different regimes. 
When $\tau_2 = \tau_3$, the model preserves $SO(2)$ with respect to the $z$ axis and conserves magnetization. 
Moreover, for $\tau_1=\tau_2=\tau_3$ the model reduces to the celebrated Heisenberg model with $SO(3)$ rotational invariance.
In the following, we investigate the higher-frequency corrections of this model, which were neglected in previous analyses.

\subsection{Effective theory: Correction $1/\omega^2$}

As detailed in Appendix~\ref{appendix_floquet} and \ref{appendix_coefficients}, the first non-vanishing correction to $H_0$ arises at the second-order in the Floquet perturbative expansion \eqref{eq: flq-exp}, and takes the form $H^{(2)}_{\text{eff}}/\omega^2 = H^{2\sigma} + H^{4\sigma}$ where $H^{2\sigma}$ and $H^{4\sigma}$ respectively correspond to two-spin and four-spin coupling terms explicitly given by
\begin{equation}
\label{eq:twospins_ham}
    H^{2\sigma} = \sum_{j> k}\left(\mathcal{J}^x_{j,k}\sigma_j^x \sigma_k^x+\mathcal{J}^y_{j,k}\sigma_j^y \sigma_k^y+\mathcal{J}^z_{j,k}\sigma_j^z \sigma_k^z\right)\,,
\end{equation}
and
\begin{align}
\label{eq:fourspin_ham}
    \nonumber H^{4\sigma} =\sum_{\substack{h>i\\ j>k }}\,^{'}
         \left(\right.&C^{yz}_{hi;jk}\sigma^y_h\sigma^y_i\sigma^z_j\sigma^z_k  + C^{zx}_{hi;jk}\sigma^z_h\sigma^z_i\sigma^x_j\sigma^x_k \\
         & +C^{xy}_{hi;jk}\sigma^x_h\sigma^x_i\sigma^y_j\sigma^y_k \left.\right)\,,
\end{align}
where the restricted summation $\sum '(\dots)$ excludes terms with $(h,i)=(j,k)$.

The Hamiltonian $H^{(2)}_{\text{eff}}$ does not possess in general any explicit relevant symmetry except for the trivial $\mathbb Z_2$ parity ($\vec{\sigma} \rightarrow - \vec{\sigma}$). 
However, under certain conditions an approximate $SO(2)$ symmetry emerges. 
In particular, when $\tau_2 = \tau_3$ and $\tau_1\ll \tau_2,\tau_3$ the two-spin and four-spin couplings satisfy $\mathcal{J}^x_{j,k}=\mathcal{J}^y_{j,k}$ and $C^{yz}_{hi;jk} = C^{zx}_{jk;hi}$. 
Explicit $SO(2)$ symmetry breaking originates exclusively from terms of the form $\sigma^x_h\sigma^x_i\sigma^y_j\sigma^y_k$ which reduce the symmetry to the $\mathbb Z_4$ subgroup of rotations by $\pi/2$ around the $z$ axis.
In the ideal-pulse limit introduced above, taking $\tau_1\to 0$ brings the pulse at the end of one period, with phase $\phi=\pi$, adjacent to the pulse at the beginning of the next period, with phase $\phi=0$. These two $\pi/2$ rotations are about opposite axes and therefore cancel each other. The protocol can thus be represented as an effective two-pulse sequence over one period. In the rest of the manuscript, we will consider a simplified WAHUHA sequence corresponding to second and third pulses with $\tau_2=\tau_3=T/4$ with a rotation around the $y$ axis.

In order to simplify our discussion, we limit ourselves for the moment to nearest-neighbor $J_{ij}$ couplings in Eq.~\eqref{eq:XY}
and we fix the lattice to be a one-dimensional chain. 
Under these assumptions, the Hamiltonian reads $H_\text{eff}\approx H_\text{eff}^{(0)}+ H_\text{eff}^{(2)}/\omega^2 = H_{SO(2)}+H_{\cancel{SO(2)}}$, where the two terms take the form
\begin{equation}
\label{eq:effective_1d_nn}
    \begin{split}
        & H_{SO(2)} = 
        \\& \; \left(1-\Gamma \right)\sum_{\substack{i}}\left(\sigma^+_i\sigma^-_{i+1}+\sigma^-_i\sigma^+_{i+1}+\sigma^z_i\sigma^z_{i+1}\right)\\
        &+\Gamma \sum_{\substack{i}}\left(\sigma^+_i\sigma^-_{i+2}+\sigma^-_i\sigma^+_{i+2}-\sigma^z_i\sigma^z_{i+2}\right)\\
        &-\Gamma \sum_i \left(\sigma^z_{i-1}\sigma^z_{i+1}+\sigma^z_{i+1}\sigma^z_{i+3}\right)\left(\sigma^+_i\sigma^-_{i+2}+\sigma^-_i\sigma^+_{i+2}\right)\\
        &+2\Gamma \sum_i \sigma^z_{i+1}\sigma^z_{i+2}\left(\sigma^+_i\sigma^-_{i+3}+\sigma^-_i\sigma^+_{i+3}\right)\\
        &+2\Gamma \sum_i\left(\sigma^+_i\sigma^+_{i+1}\sigma^-_{i+2}\sigma^-_{i+3} + \sigma^-_i\sigma^-_{i+1}\sigma^+_{i+2}\sigma^+_{i+3}\right),
\end{split}
\end{equation}
and
\begin{equation}
\label{eq:symm-break-terms}        H_{\cancel{SO(2)}} = -2\Gamma \sum_i\sigma^+_i\sigma^+_{i+1}\sigma^+_{i+2}\sigma^+_{i+3} + \text{h.c.},
\end{equation}
assuming units $J_{\text{nn}}\equiv J_{i,i+1}=1$. 
The reduction from the general second-order Hamiltonian to Eqs.~\eqref{eq:effective_1d_nn} and~\eqref{eq:symm-break-terms} is summarized in App.~\ref{app:nn_reduction}.
As it is evident from the equations above, $H_{SO(2)}$ contains terms all of which conserve the total magnetization, while $H_{\cancel{SO(2)}}$ includes a term violating this conservation law. 
Within Floquet perturbation theory we can readily obtain the coefficient $\Gamma\equiv \frac{1}{6}\left(\frac{J_{\text{nn}} \pi}{\omega}\right)^2$, as shown in App.~\ref{appendix_coefficients}.

The symmetric Hamiltonian in Eq.~\eqref{eq:effective_1d_nn} can be interpreted as a sum of distinct physical contributions. 
The first line represents the leading-order Hamiltonian, which is a nearest-neighbor XXZ model and includes higher-order Floquet corrections via $\Gamma$. 
The second line introduces longer-range corrections to the previous XXZ model proportional to $\Gamma$, including next-nearest-neighbor hopping and interactions.
Starting from the third line, higher-order spin correlations appear. The third and fourth line describe correlated hopping processes whereas
the fifth and last line corresponds to pair-hopping processes, in which two spins hop together across adjacent sites, effectively shifting a pair by two sites. 
Crucially, the emergent interactions described by the operator $\sigma^z_{i+1}\sigma^z_{i+2}\left(\sigma^+_i\sigma^-_{i+3}+\sigma^-_i\sigma^+_{i+3}\right)$ are relevant for the mobility of trimers, as we will further discuss below. 
In contrast, the term in Eq.~\eqref{eq:symm-break-terms} changes the magnetization by four units and can therefore couple the trimer sector to higher-magnetization sectors. In the following dynamical analysis, we monitor whether this channel obscures the trimer signal on the simulated timescales.
The resulting Floquet Hamiltonian dynamics has been benchmarked with the full exact dynamics in Appendix~\ref{appendix: benchmark}. The benchmark shows that the second-order effective Hamiltonian captures the local magnetization dynamics more accurately than the leading-order XXZ model in the parameter regimes considered, while global quantities such as the total magnetization can display visible deviations in larger systems due to higher-order Floquet corrections.

\section{Trimer dynamics}
\label{sec:trimer_dynamics}

In this section we investigate the dynamics of three–body bound states (hereafter referred to as trimers) in the effective models derived above. 
In the static nearest–neighbor XXZ chain such excitations are known to be very heavy, as their motion arises only through higher–order virtual processes.  
The Floquet-engineered Hamiltonian introduced in Sec.~\ref{sec:WAHUHA} instead contains additional hopping and correlated terms whose contribution to trimer motion becomes explicit in the simplified one-dimensional, nearest–neighbor setting adopted here.  
We first recall the structure and dispersion of the trimer in the XXZ limit, and then analyze how Floquet corrections alter its mobility and its robustness against residual $SO(2)$–breaking channels.

\subsection{What is a trimer?}

We first consider the nearest-neighbor spin-1/2 XXZ Hamiltonian on a generic lattice geometry,
\begin{equation}
\label{eq:xxz_model_generic}
    H =
    2h \sum_{\langle i,j\rangle}
    \left(
    \sigma^+_i \sigma^-_j
    +
    \sigma^-_i \sigma^+_j
    \right)
    +
    \Delta \sum_{\langle i,j\rangle}
    \sigma^z_i \sigma^z_j ,
\end{equation}
where $\langle i,j\rangle$ denotes pairs of nearest-neighbor sites.

In one dimension with closed boundary conditions, Eq.~\eqref{eq:xxz_model_generic} reduces to
\begin{equation}
\label{eq:xxz_model}
    H = 2h \sum_i \left( \sigma^+_i \sigma^-_{i+1} + \sigma^-_i \sigma^+_{i+1} \right) + \Delta \sum_i \sigma^z_i \sigma^z_{i+1}.
\end{equation}
We focus on the ordered phase regime $|h/\Delta| < 1$ and $\Delta > 0$, where the Ising interaction dominates.
The leading order term in the effective Hamiltonian \eqref{eq:effective_1d_nn} corresponds to the XXZ model for $h/\Delta=0.5$.

For the purpose of studying the dynamics of spin-flip clusters, it is convenient to adopt the fully polarized state $ |0\rangle = |\downarrow \downarrow \dots \downarrow\rangle $ as a reference. 
Although this state corresponds to the highest energy configuration in the antiferromagnetic regime, it provides a simple and intuitive background in which flipped spins can propagate, interact, and form bound states. 
Within this picture, a single flipped spin $|i\rangle_S\equiv|\dots \downarrow \downarrow \uparrow_{i} \downarrow \downarrow \dots \rangle$ (i.e. a \emph{magnon}) plays the role of a mobile particle, enabling a clean description of its dynamics.
Two magnons can form bound states $|i\rangle_D\equiv|\dots \downarrow \uparrow_{i} \uparrow_{i+1} \downarrow \dots \rangle$, here dubbed dimers, with energy different from free magnons. 
The XXZ model also supports multiparticle bound states \cite{ganahl_2012, Gross2013, kranzl_2023, kim_2024} and here we consider the case of three adjacent spin flips: a ``trimer'' \mbox{$|i\rangle_T = |\dots \downarrow \uparrow_i \uparrow_{i+1} \uparrow_{i+2} \downarrow \dots \rangle$}.

The existence and stability of trimers can be understood perturbatively by considering the subspace of states $|i\rangle_T$ weakly coupled by the spin exchange term proportional to $h$.
The effective low-energy Hamiltonian takes the form of a tight-binding model:
\begin{equation}
    \label{eq:tb_trimer}
    \mathcal{H}_T = \sum_i \epsilon_T \ket{i}\bra{i} + t_T \left( \ket{i+1}\bra{i} + \text{h.c.} \right) ,
\end{equation}
where $\ket{i}\equiv\ket{i}_T$ denotes the trimer localized at sites $i, i+1, i+2$, $\epsilon_T$ is the on-site energy of the trimer state, $t_T$ is the effective hopping amplitude. From perturbation theory, one finds:
\begin{equation}
    \epsilon_T = E_0 + 2\frac{h^2}{\Delta} + \mathcal{O}(h^4),
\end{equation}
where $E_0 = (L-4)\Delta$ with $L$ the number of sites of the chain with closed boundary conditions.
The trimer hopping amplitude reads:
\begin{equation}
t_T = \frac{h^3}{2\Delta^2}\,,
\end{equation}
where the only relevant virtual states are spin patterns of the form $\ket{\dots\downarrow\uparrow\downarrow\uparrow\uparrow\downarrow\dots}$ and $\ket{\dots\downarrow\uparrow\uparrow\downarrow\uparrow\downarrow\dots}$, whose energy difference from the trimer manifold is $ 4\Delta $. 
From Eq.~\eqref{eq:tb_trimer}, the trimer dispersion relation reads:
\begin{equation}
    E_T(k) =  E_0 +2\frac{h^2}{\Delta} + \frac{h^3}{\Delta^2}\cos k.
\end{equation}
It is important to emphasize that the trimer band described above represents only a subset of the full spectrum at fixed total magnetization $ N_\uparrow = 3 $, which is represented in Fig.~\ref{fig:bands}(a, b). 
In this sector, one also finds states where the three spin-ups are not adjacent, such as configurations with two nearby spins and one separated (i.e. those used above for enabling the trimer mobility), or three completely separated spins. 
These configurations give rise to additional bands in the energy spectrum yielding bound state bands and scattering continua.

From a perturbative perspective, the coherent propagation of the trimer is well-defined as long as its band remains separated from the scattering states, namely for sufficiently small values of $h$. 
As $ h $ increases, the trimer band lowers in energy and eventually overlaps with the scattering continuum triggering conversion of trimers into lower-spin constituents (\emph{i.e.} magnons and 2-magnon bound states) 

A crucial ingredient for identifying trimers as coherent quasiparticles is the conservation of the total magnetization, i.e. an $SO(2)$ symmetry. As visible from the band structure in Fig.~\ref{fig:bands}(c), the trimer band can overlap in energy to single-magnon states and other string-like excitations that share the same number of domain walls but differ in spin number, such as dimer bound states. Without magnetization conservation, such resonances would enable mixing between these sectors, obscuring the identification of trimers as coherent, long-lived modes.

Finally, it is worth noting that trimers move significantly much slower than other excitations. 
Their effective hopping amplitude arises at third order in perturbation theory via two virtual states, scaling as $ t_T \sim h^3 / \Delta^2 $. 
In contrast, a single magnon propagates directly at first order in $ h $, while a two-magnons bound state hops at second order with amplitude $ \sim h^2 / \Delta $. 
As a result, the trimer band appears almost flat compared to the others (Fig.~\ref{fig:bands}(c)), and its group velocity is suppressed. 
This slow dynamics can be a limiting factor for experimental detection, since the trimer motion may occur over timescales longer than the coherence times of available platforms. 
This motivates the effort to identify schemes that allow to enhance its dynamics and visibility through suitable engineering of system parameters, as we discuss in this work and more specifically in the next paragraph.

\subsection{Trimer dynamics and SO(2) leakage}
\label{subsec:trimer_large_systems} 

\begin{figure}
    \centering
    \includegraphics[width=\linewidth]{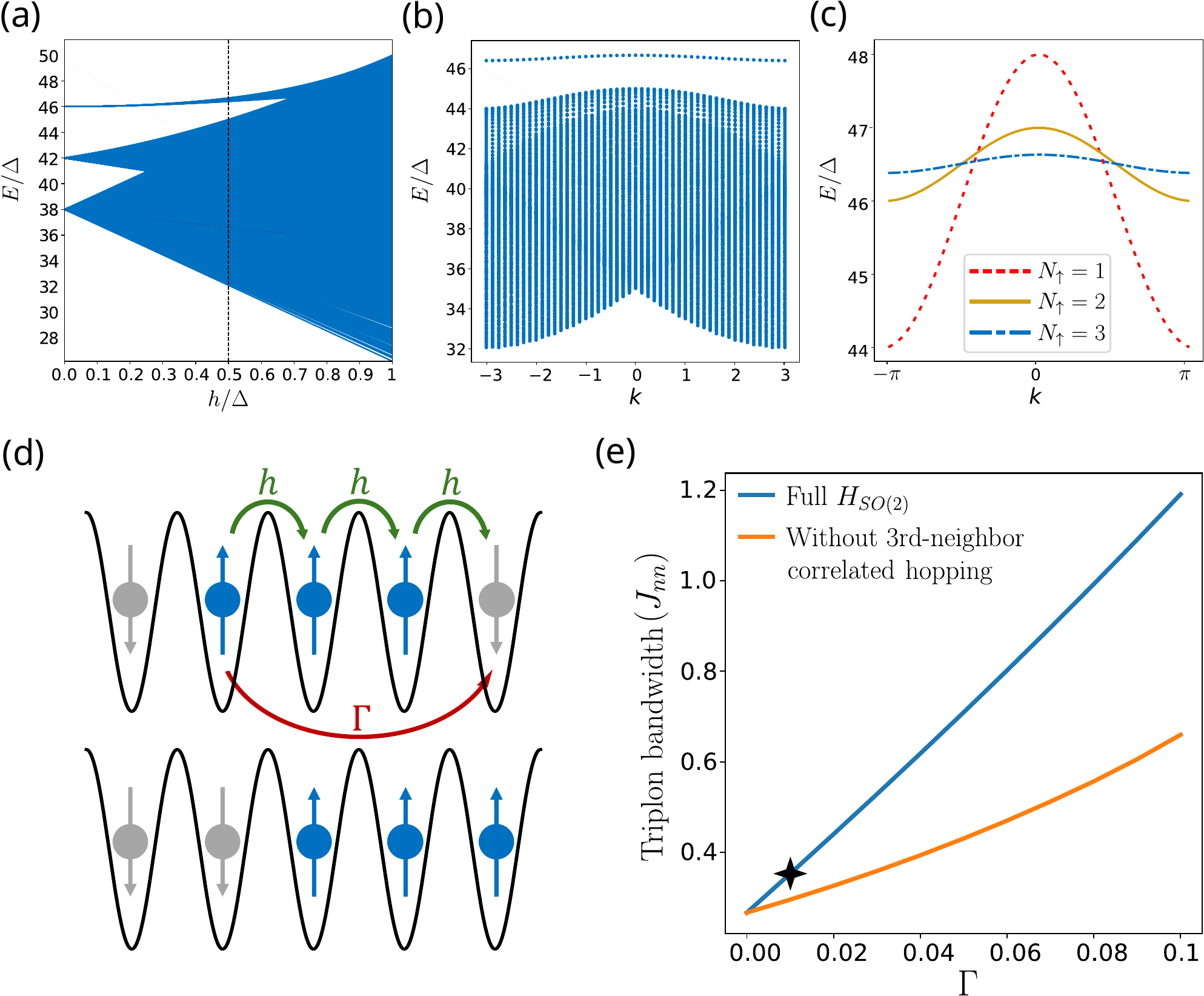}
   \caption{
\justifying
Trimer properties in the XXZ model and its Floquet corrections. 
\textbf{(a)} Energy spectrum of the 1D XXZ model (with closed boundary conditions) in the $N_\uparrow = 3$ sector for $L = 50$ sites as a function of $h/\Delta$. 
The trimer band appears as a narrow, high-energy band separated from lower-energy scattering states for $h/\Delta \lesssim 2/3$. 
The vertical dashed black line marks the reference value $h/\Delta = 0.5$, corresponding to the value for the leading order XXZ terms in Eq. \eqref{eq:effective_1d_nn}.
\textbf{(b)} Momentum-resolved spectrum of the $N_\uparrow = 3$ sector for $h/\Delta = 0.5$. The upper isolated branch corresponds to the trimer band discussed in the main text.
\textbf{(c)} Comparison of the highest-energy bands of the XXZ model at $h/\Delta = 0.5$ across different magnetization sectors ($N_\uparrow = 1$, 2, and 3) highlighting the different dispersion of magnons, dimers and trimers. 
\textbf{(d)} Schematic illustration of the correlated third-neighbor hopping process contributing to trimer propagation in the effective Hamiltonian. 
\textbf{(e)} Bandwidth of the trimer band from Eq.~\eqref{eq:effective_1d_nn} as a function of the Floquet parameter $\Gamma$ for the full effective Hamiltonian (blue line) and the reduced model (orange line) where the correlated third-neighbor hopping term (fourth line in Eq.~\eqref{eq:effective_1d_nn}) is removed.
The black star marks the choice of $\Gamma = 0.01$, used throughout the main text.
}
    \label{fig:bands}
\end{figure}

Perturbations introducing correlated longer-range hopping processes can significantly enhance the trimer mobility, thereby speeding up the propagation timescales. A concrete example of such a modification is provided by the effective model previously introduced in Eq.~\eqref{eq:effective_1d_nn}, where a small parameter $ \Gamma \ll 1 $ controls various next-nearest and higher-order spin interactions. 
In particular, the fourth line of Eq.~\eqref{eq:effective_1d_nn} contains a correlated third-neighbor hopping term of the form  $ \Gamma \,\sigma^z_{i+1}\sigma^z_{i+2} \left(\sigma^+_i\sigma^-_{i+3} + \sigma^-_i\sigma^+_{i+3}\right) $. 
This term directly enables a first-order coherent hopping process for the trimer as a whole, without the need of virtual intermediate steps (see Fig~\ref{fig:bands}(d)). 
However such process will interfere with the  ones originating from the rest of the Hamiltonian, in particular the leading order XXZ terms, and it is not guaranteed that it would enhance the mobility. 
Since $\Gamma>0$ and the product $ \sigma^z_{i+1}\sigma^z_{i+2} $ is positive on trimer states, the overall sign of the process is also positive. 
As shown in the previous paragraph for the perturbative regime and in Fig.~\ref{fig:bands}(a,b,c), XXZ trimers have an inverted band with positive hopping.
Thus the correlated hopping process adds up constructively to them and leads to an enhancement of the mobility rather than a suppression.
As a consequence, the effective group velocity of the trimer increases and the corresponding energy band becomes more dispersive. 
This effect is illustrated in Fig.~\ref{fig:bands}(e), which compares the trimer bandwidth for increasing values of $\Gamma$.
The presence of the third-neighbor correlated hopping in Eq.~\eqref{eq:effective_1d_nn} visibly increases the bandwidth of the trimer.
To isolate the role of this process from the other $\Gamma$-dependent corrections, the same figure also compares the full effective Hamiltonian with a reduced model in which only the correlated third-neighbor hopping term in the fourth line of Eq.~\eqref{eq:effective_1d_nn} is removed. The reduced bandwidth obtained in this case shows that this term gives the dominant contribution to the Floquet-induced enhancement of trimer mobility, while the remaining $\Gamma$-dependent terms mainly provide a smaller background renormalization.

To complement the band structure analysis, we study trimer dynamics of the XXZ Hamiltonian under the WAHUHA protocol discussed above in chains of $L = 17$ sites. 

We find that trimer propagation remains well-defined under the exact driven dynamics for $\Gamma \lesssim 0.01$ ($\omega/J_{\text{nn}}\gtrsim 8$). 
The second-order effective Hamiltonian \eqref{eq:effective_1d_nn} provides the reference bound-state dispersion used to estimate the propagation velocity, but the robustness of the signal is assessed directly from the driven time evolution.

In this regime, we observe an enhancement of the trimer propagation velocity of approximately up to 30\% compared to the undriven $\Gamma \rightarrow 0$ case.
Despite the presence of $SO(2)$-breaking terms, the dominating fast dynamics is dictated by the trimer light cone as we comment further below.

\begin{figure}[!t]
    \centering    \includegraphics[width=\linewidth]{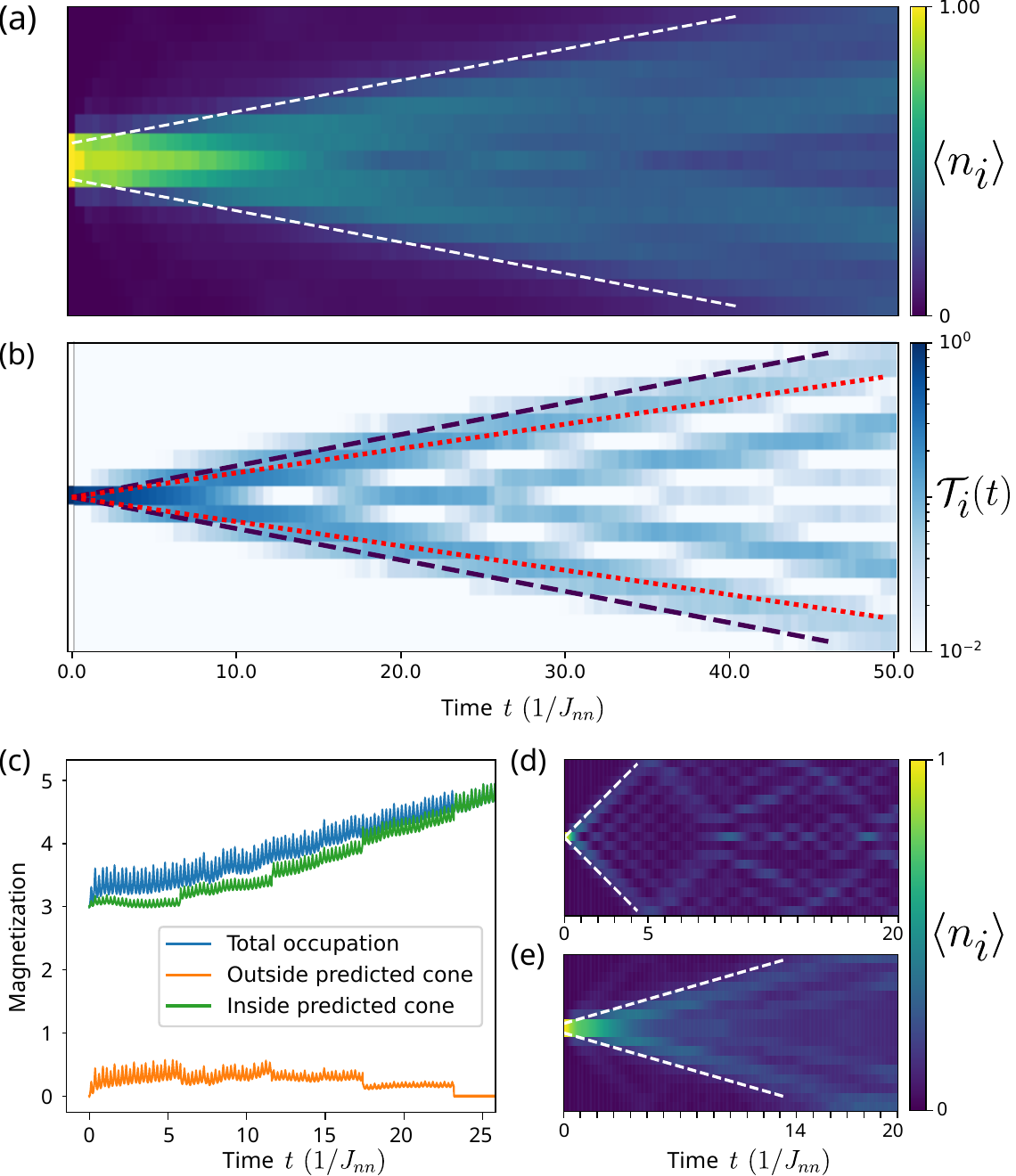}
    \caption{
    \justifying
    Time-resolved diagnostics of trimer propagation in a 17-site chain under the exact driven Hamiltonian with $\Gamma = 0.01$ and closed boundary conditions. 
    \textbf{(a)} Local magnetization as a function of time, showing a clear light-cone structure emerging from a central trimer excitation. The effective light-cone (white dashed line) is obtained from the bound-state dispersion by extracting the maximum slope via finite differences over three adjacent $k$-points in the exact spectrum. 
    \textbf{(b)} Time evolution of the trimer correlator $\mathcal{T}_i(t)$ showing ballistic behavior.
    The dashed dark line marks the effective light-cone velocity for $\Gamma = 0.01$ ($v \approx 0.173$), while the dotted red line shows the velocity for $\Gamma \to 0$ ($v \approx 0.130$) for comparison.
    \textbf{(c)} Time evolution of the magnetization $m=\sum_i n_i$ inside the light cone (green), outside of the cone (orange) and the total one (blue). The step-like structure of the curves arises the discrete nature of the lattice. \textbf{(d)} Magnetization profile of a propagating single-spin magnon and \textbf{(e)} of a dimer.
    }
\label{fig:trimer_diagnostics_L17}
\end{figure}

We support the above conclusions by monitoring the quantum walk dynamics of an initial product state \mbox{$|i\rangle_T = |\dots \downarrow \uparrow_i \uparrow_{i+1} \uparrow_{i+2} \downarrow \dots \rangle$} using three complementary observables. 
First, we monitor the time evolution of the local magnetization, $n_i = (1+\sigma^z_i)/2$, which reveals a coherent ballistic front that is consistent with the light cone velocity $v_
\text{max}$ obtained from the effective theory  Eq.~\eqref{eq:effective_1d_nn}, see Fig.~\ref{fig:trimer_diagnostics_L17}(a). 
Second, we compute a 5-point trimer correlator, specifically designed to detect isolated three-spin bound states surrounded by spin-down spins \cite{preiss_strongly_nodate, rapp_ultracold_2012}: 
\begin{equation}
\label{eq:trimer_correlator}
\mathcal{T}_i(t) = \left\langle (1-n_{i-2})n_{i-1}n_{i}n_{i+1}(1-n_{i+2})\right\rangle\,.   
\end{equation}
This correlator clearly traces the same light-cone structure observed in the local magnetization, see Fig.~\ref{fig:trimer_diagnostics_L17}(b).
Finally, we quantify the total magnetization contained inside and outside the light-cone. 
We find that the magnetization outside the light-cone remains small and approximately constant while the magnetization inside the cone increase with time.
This indicates that processes not conserving magnetization are dominantly slower than the trimer propagation and do not hinder the identification of its dynamics.
Altogether, these observations show that, on the simulated timescales, the magnetization-nonconserving channels generated by the drive do not prevent the identification of the trimer propagation front for $\Gamma \lesssim 0.015$.

For comparison, we also simulate the dynamics of a single magnon and a two-spin dimer under the same driven Hamiltonian, using initial states respectively given by $|\dots \downarrow \uparrow_i \downarrow \dots \rangle$ and $|\dots \downarrow \uparrow_i \uparrow_{i+1} \downarrow \dots \rangle$. In both cases we employ the same system size ($L=17$), drive strength ($\Gamma = 0.01$), and closed boundary conditions used in the trimer case. The corresponding magnetization profiles, shown in Fig.~\ref{fig:trimer_diagnostics_L17}(d,e), illustrate the faster propagation of these excitations compared to the trimer.
We checked that the dynamics with closed boundary conditions is reproduced also for an open chain with a sufficiently large number of sites $L\ge 50$ to minimize boundary effects. 

\section{Long-range and higher-dimensional effects}
\label{sec:LongRange2D}

In the previous sections, we have focused on trimer dynamics in one-dimensional systems with nearest-neighbor interactions. While this setup captures the essential physical mechanisms and allows for clear analytical treatment, it is important to address generalizations that bring the model closer to experimental platforms and open further directions. Here we consider two such extensions: the inclusion of long-range interactions and the generalization to two-dimensional lattice geometries.

\subsection{Long-range interactions}
\label{sec:long-range}
One of the most natural and experimentally relevant extensions concerns systems with long-range interactions as compared to the nearest-neighbor couplings considered above. 
In particular, platforms based on Rydberg atoms trapped in optical tweezers provide a natural setting for realizing spin models with tunable long-range dipolar interactions \cite{browaeys_many-body_2020, de_leseleuc_observation_2019, de_leseleuc_optical_2017}, while alternative realizations include magnetic dipolar atoms \cite{chomaz2022dipolar} and polar molecules \cite{Zoller2006, Hazzard2024,lu2026probing}. 
Rydberg systems implement an XY Hamiltonian, Eq.~\eqref{eq:XY}, through dipole-dipole exchange processes, and an external magnetic field determines the angular structure of the interactions. 
Specifically the atom-atom interactions in Eq.~\eqref{eq:XY} scale as $J_{ij} = C_{dd} (1 - 3 \cos^2 \theta_{ij})/r_{ij}^3 $, with $ \theta_{ij} $ the angle between the magnetic field and the interatomic axis \cite{ravets_measurement_2015}. 
This provides a convenient knob to tune anisotropy in the atom-atom couplings as well as the sign.
In this work, we consider lattice geometries (e.g. a one-dimensional chain or a two-dimensional lattice) for which we can choose the angle $\theta_{ij}=\pi/2$ for all pairs of atoms $i$ and $j$ and  obtain isotropic dipolar interactions decaying as $1/r_{ij}^3$. 
More generally, one can however ask how the results of this work depend on the decay of the interactions with distance as $1/r^\alpha$, with $\alpha=3$ being the dipolar case. 
This is motivated by the fact that properties of bound states can indeed strongly depend on the long-range tail \cite{macri_bound_2021, Tecer2024} and different platforms provide distinct and possibly tunable long-range couplings \cite{Defenu2023}. 

\begin{figure}
    \centering
    \includegraphics[width=\linewidth]{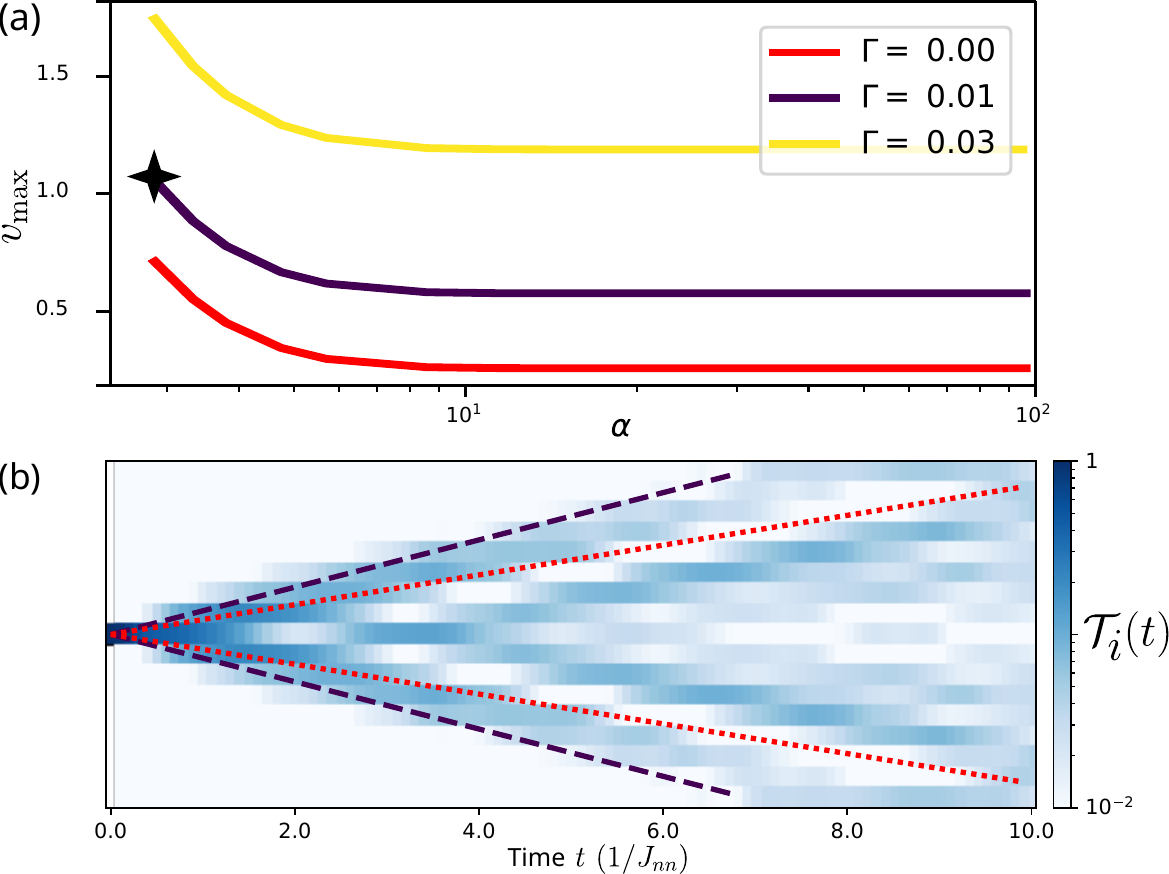}
    \caption{\justifying
    \textbf{(a)} Maximum trimer velocity $v_{\text{max}} = \frac{\partial E_T(k)}{\partial k}$ evaluated from the exact-diagonalization spectrum of the $H_\text{eff}$ up to $\mathcal O(1/\omega^2)$ specializing Eqs.~\eqref{eq:twospins_ham}, \eqref{eq:fourspin_ham} to long-range couplings $J_{ij}\sim 1/r_{ij}^\alpha$, $L=17$ sites with closed boundary conditions, $\tau_1=0$ and $\tau_2=\tau_3=T/4$, within the sector $N_\uparrow=3$, thus projecting out all magnetization non-conserving terms. 
    The distance $r_{ij}$ is taken as the minimal distance on the periodic chain.
    The lower cutoff $\alpha = 3$ corresponds to the dipolar case. The black star marks the choice of $\Gamma$ and $\alpha$ used in section \ref{sec:long-range}.
    \textbf{(b)} Time evolution under the exact driven Hamiltonian of the trimer projection operator $\mathcal{T}_i(t)$ for $\Gamma = 0.01$ and $\alpha = 3$ (dipolar couplings). We observe an increase of $\sim60\%$ in the speed of the excitations with respect to the $\Gamma\rightarrow 0$ case (red dotted line).}
    \label{fig:longrange}
\end{figure}

The Floquet analysis presented in previous sections can be extended in the presence of long-range couplings, leading to an effective model analogous to Eq.~\eqref{eq:effective_1d_nn} but containing additional long-range interaction channels, including new terms that break $SO(2)$ (see Eq.~\ref{eq:general_hamiltonian}). 
To probe the impact of these additional interactions on trimer dynamics we computed the maximum group velocity of the trimer band for various values of $\alpha$ and Floquet strengths $\Gamma$ in the sector $N_\uparrow=3$ (shown in Fig.~\ref{fig:longrange}(a)), thus restricting the analysis to the $SO(2)$ symmetric terms. 
We find that the trimer mobility substantially increases as $\alpha$ decreases, \emph{i.e.} as interactions become longer-ranged, while the dynamics remains effectively bounded by a linear light cone~\cite{FossFeig2015}. 
Not only does the natural (zeroth-order) trimer velocity increases, but the additional enhancement due to second-order corrections is also amplified at low $\alpha$ values.
This analysis is supported by real-time dynamics showing that the predicted velocity corresponds to an analogous propagation cone, see Fig.~\ref{fig:longrange}(b) for $\alpha=3$.
Specifically, we find that the Floquet drive induces an enhancement up to 60\% faster than the undriven propagation. 
Furthermore, a comparison with the nearest-neighbor case studied in Fig.~\ref{fig:trimer_diagnostics_L17}\edo{(b)} shows a clear drastic improvement of the propagation time scale. 
Here the trimer reaches the edge after 6 tunneling times $J_{\text{nn}}$, whereas in the nearest-neighbor case it was a much slower process requiring 45 tunneling times for a chain with 17 sites. 

\begin{figure}
    \centering
    \includegraphics[width=\linewidth]{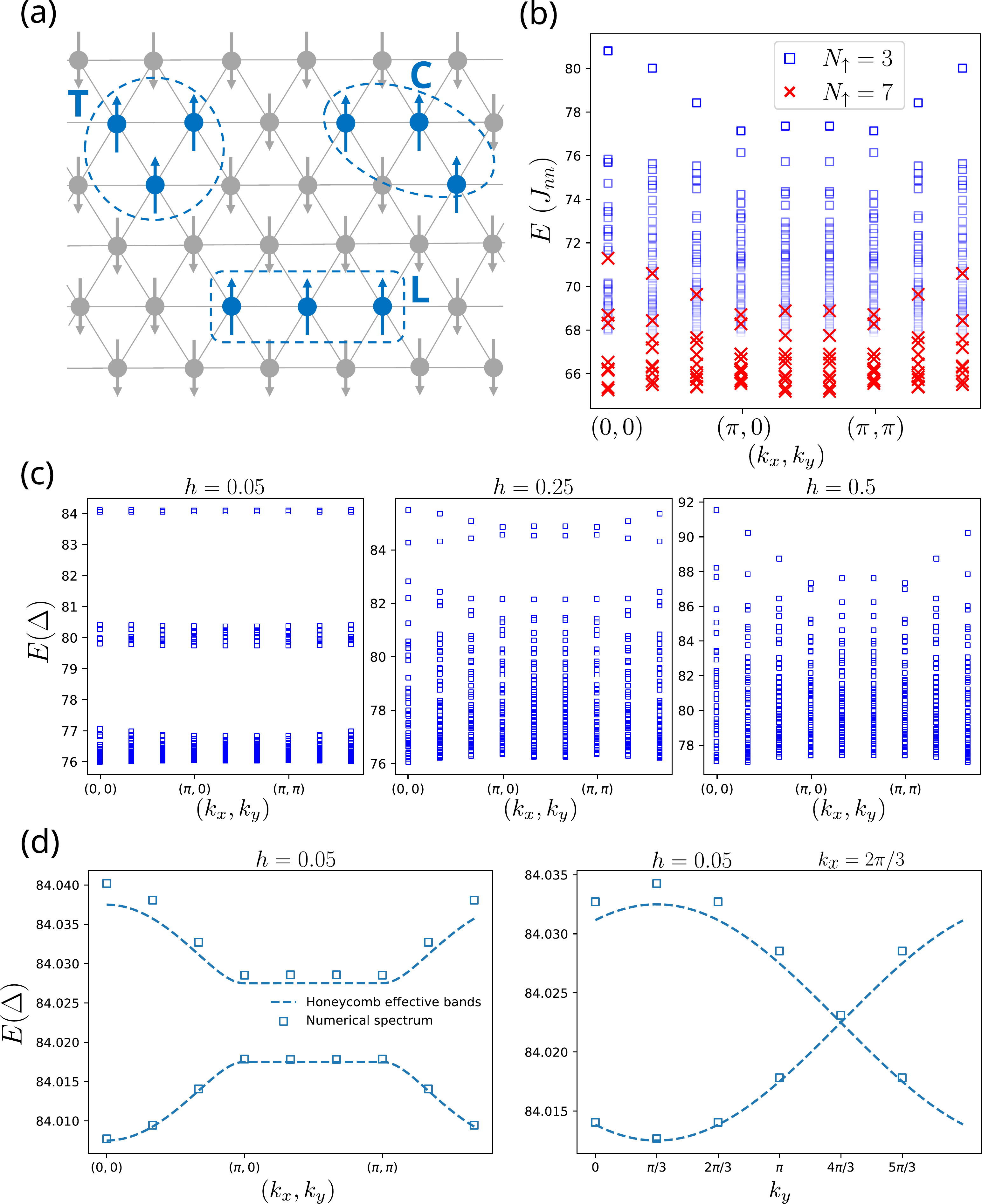}
    \caption{ 
    \justifying
    Trimer bound states in a 2D triangular lattice.
    \textbf{(a)} Schematic representation of compact three-spin configurations on a triangular lattice. 
    \textbf{(b)} Spectrum of the effective Hamiltonian~(see Eq.~\ref{eq:general_hamiltonian} in the Appendices) with $\Gamma = 0.015$ on a 2D triangular lattice of size $L\times L$ sites with $L=6$, nearest-neighbor couplings and closed boundary conditions along both directions. The plot shows energy levels in the $N_\uparrow = 3$ sector (blue squares) and the $N_\uparrow = 7$ sector (red crosses) along a momentum-space path in the 2D Brillouin zone.
    The highest-energy branch in the $N_\uparrow = 3$ sector appears isolated from both lower-energy $N_\uparrow = 3$ excitations and higher-spin sectors such as $N_\uparrow = 7$.
    \textbf{(c)} Spectrum of the triangular-lattice XXZ model of Eq.~\eqref{eq:xxz_model_generic} in the $N_\uparrow=3$ sector for $\Delta=1$ and increasing values of $h$. In the small-$h$ limit, the highest-energy states are identified with compact three-spin configurations. Their continuous evolution with increasing $h$ supports the trimer-like interpretation of the isolated high-energy branch in panel (b). 
    \textbf{(d)} Comparison between the exact spectrum of T states and their effective honeycomb-lattice description (see main text) for different cuts of the Brillouin zone.
    }
    \label{fig:triangularlattice}
\end{figure}

Recent experiments demonstrate that the characteristic interaction strengths and timescales assumed in this work are compatible with state-of-the-art Rydberg quantum simulators. 
For example,  in the Rydberg implementation by Scholl et al. \cite{scholl_microwave-engineering_2022} and by Chen et al. 
\cite{chen_spectroscopy_2023}
dynamics up to a few tunneling time ($\hbar/J_{\text{nn}}$) is achieved, which is within the range required by our findings as in Fig.~\ref{fig:longrange}.

\subsection{Two-dimensional geometries}

We now briefly discuss the extension of our analysis to two-dimensional lattice geometries, where \emph{i)} trimer states can further gain additional mobility due to the higher-dimensional geometry and \emph{ii)} energetic isolation of trimer states can be observed thus providing a route towards the suppression of resonant processes into other magnetizazion sectors that were identified in 1D.

To be specific we focus on a triangular geometry with nearest-neighbor couplings, as depicted in Fig.~\ref{fig:triangularlattice}(a), but other geometries can provide similar mechanisms with distinct features.
As shown in the figure, there are different kinds of compact three-spin configurations, distinguished by their geometric structure, that can have 12 or 14 domain walls.
This provides an energy cost $\Delta E = 24-28J$ with respect to the fully polarized state.
Instead, the $7$-spin excitations have 18-30 domain walls with an energy cost $\Delta E = 36-60J$.
The compact three-spin configurations are of three types: triangles (T) and linear (L) or chair-shaped (C) states, shown in Fig.~\ref{fig:triangularlattice}(a).
Notice that the T configurations are not degenerate with the L and C configurations due to the different count of domain walls. 

In Fig.~\ref{fig:triangularlattice}(b), we show the ED spectrum of the Hamiltonian in Eq.~(\ref{eq:general_hamiltonian}) of the Appendices for a small lattice $6\times6$, where the two sectors $N_\uparrow = 3$ and $N_\uparrow = 7$ are plotted for $\Gamma = 0.015$.

We find an isolated high-energy branch in the $N_\uparrow=3$ sector. 
To clarify the nature of this branch, Fig.~\ref{fig:triangularlattice}(c) shows the evolution of the corresponding momentum-resolved spectrum of the triangular-lattice XXZ model of Eq.~\eqref{eq:xxz_model_generic} in the $N_\uparrow=3$ sector as $h$ changes. 
In the small-$h$ limit, where T, C and L states are approximately eigenstates of the model, the highest-energy states can be directly associated with compact T configurations through the domain-wall counting discussed above. 
Notice in particular that there are two types of T states, depending whether the triangle points up or down. 
This gives rise to two distinct T bands.
Since we consider a system of $N=L\times L=36$ sites with $L=6$ and periodic boundary conditions along both axes, the energy of the fully polarized (FP) state is given by the total number of nearest-neighbor links, namely $N_\ell=3 L^2 =108$, thus $E_{FP}=108\Delta$. 
T configurations have 12 domain walls corresponding to energy $\Delta E_T= E_{FP}-E_T=108\Delta -2\times12\Delta=84\Delta$. 
C and L configurations have 14 domain walls and thus $\Delta E_C=\Delta E_L = 80 \Delta$.
These energies are clearly identified in the left panel of Fig.~\ref{fig:triangularlattice}(c). 
Notice that 
As $h$ is increased, the different energy bands gain dispersion and bandwidth, so they start overlapping and merging. However, the upper band remains isolated and gapped from the rest. 
This supports the identification in Fig.~\ref{fig:triangularlattice}(b) of the upper state as a trimer band, in close analogy with the one-dimensional trimer band discussed in Sec.~\ref{subsec:trimer_large_systems} and shown in Fig.~\ref{fig:bands}(b).

The comparison with the $N_\uparrow=7$ sector in Fig.~\ref{fig:triangularlattice}(b) further shows that, for the parameters considered here, this trimer branch is energetically separated from it, thus suppressing the coupling originating from leading $SO(2)$-breaking process. Moreover, compared with the zeroth-order XXZ spectrum shown in Fig.~\ref{fig:triangularlattice}(c), the inclusion of the second-order Floquet terms in Fig.~\ref{fig:triangularlattice}(b) appears to further gap and isolate the upper branch from the lower $N_\uparrow=3$ states. This suggests that the Floquet-induced corrections can contribute not only to the mobility of compact three-spin excitations, but also to their spectral separation in triangular geometries.
A full time-dependent analysis in 2D however requires higher computational resources or strategies compared to those discussed so far, and is thus beyond the scope of this work.

The value of $\Gamma$ used in Fig.~\ref{fig:triangularlattice}(b) is larger than the one employed in the one-dimensional dynamical simulations. Increasing $\Gamma$ enhances the second-order processes discussed in this work and, at the same time, is expected to make higher-order Floquet corrections progressively more relevant. The spectral separation observed in the triangular geometry therefore suggests a possible route to protect compact three-spin states even in regimes where additional magnetization-changing channels may become more important.
Two-dimensional triangular arrays are also directly relevant for current experimental platforms \cite{scholl_quantum_2021, qiao2025kinetically}, providing a natural setting where such features could be further investigated.

Moreover, the multiband structure of compact three-spin excitations in lattices different from one-dimensional chains offers an opportunity to investigate non-trivial bound state dynamics with emergent topological properties~\cite{Salerno2020, giudice_2022}.
In particular, following Ref.~\cite{Salerno2020} and focusing for simplicity on the regime of tightly-bound T states, their center of mass sits on the center of the triangular plaquettes.
This dual lattice has the geometry of a honeycomb lattice.
At leading order, T states in nearest-neighbor plaquettes are coupled via second order processes with hopping $J_\mathrm{eff}=2h^2/\Delta$ and onsite energy $E_\mathrm{eff}=84\Delta+9h^2/\Delta$. We thus expect the bands of T states to be graphene-like, namely two bands touching through Dirac cones. 
Indication of this physics and band structure can be identified in the two upper bands of Fig.~\ref{fig:triangularlattice}(c), limited by finite size resolution, that are shown in Fig.~\ref{fig:triangularlattice}(d). 
Increasing values of $h$ lead to a partial overlap of these bands with the other bound states and with the scattering continua.

\section{Conclusions}
\label{sec:conclusions}

In this work, we have explored the emergence and dynamics of trimer bound states in periodically driven spin models engineered via Rydberg atom arrays. 
By applying a WAHUHA-type Floquet sequence to an underlying dipolar XY Hamiltonian, we obtained an effective XYZ model with tunable anisotropies and second-order corrections that include correlated hopping and multispin interactions.

Our analytical and numerical results demonstrate that second-order Floquet terms enhance trimer mobility in a regime of the protocol that approximately conserves magnetization, enabling longer-range correlated motion. 
Despite the presence of terms breaking magnetization conservation in the Floquet second-order expansion, we find that trimer propagation remains clearly identifiable in the parameter regimes considered. The second-order effective Hamiltonian quantitatively captures the driven dynamics in small systems and provides a useful description of the dominant local propagation features in larger chains.
We further find that long-range interactions enhance the propagation of trimer bound states by increasing their effective mobility.
In addition, our analysis of two-dimensional triangular geometries reveals isolated high-energy branches in the $N_\uparrow=3$ sector, suggesting the possible emergence of energetically isolated trimer-like bound states beyond one dimension.

These findings provide a new route to engineer and study multiparticle bound states in programmable quantum simulators. 
The parameter regimes considered are directly relevant to current experimental platforms based on Rydberg atoms, where Floquet periods and interaction scales naturally fall in the range of few microseconds.  
Furthermore, recent advances with polar molecules also provide an alternative setting for our findings~\cite{Ye2024}.
Our work provides further support for beyond-leading-order analyses of Floquet-engineered Rydberg systems as a tool to investigate non-equilibrium dynamics of many-body quantum states.
An interesting perspective is to analyze how alternative pulse sequences~\cite{Zhou2024} yield interaction channels distinct from the ones analyzed here, which can significantly affect the mobility of multiparticle states.

\section{Acknowledgments}
This work has received funding under the Horizon Europe programme HORIZON-CL4-2022-QUANTUM-02-SGA via the project 101113690 (PASQuanS2.1), the European Union-NextGenerationEU within the National Center for HPC, Big Data and Quantum Computing (Project No. CN00000013, CN1 Spoke 10: ‘Quantum Computing’), and has been supported by the INFN project Iniziativa Specifica IS-Quantum, the Italian Ministry of University and Research via the Excellence grant 2023-2027 ``Quantum Frontiers",  the Rita Levi-Montalcini program. 
L.M. has been supported by the project "BALANCY" under the MSCA Seal of Excellence @Unipd programme.  

\appendix

\section{Details of Floquet}
\label{appendix_floquet}

We consider a generic time-dependent Hamiltonian composed of a static and a periodic contribution,
\begin{equation}
    H'(t) = H_0 + W'(t),
\end{equation}
where $H_0$ is time independent and $W'(t)$ is a periodic operator with period $T = 2\pi/\omega$.
The periodic part can be expanded in a Fourier series as
\begin{equation}
    W'(t) = \sum_{n \geq 1} \left( V^{(n)} e^{i n \omega t} + V^{(n)\dagger} e^{-i n \omega t} \right).
\end{equation}

As a concrete example, we consider the rotated Hamiltonian $H'(t)$ defined in Eq.~(\ref{eq:rotated_ham}). 
This Hamiltonian can be decomposed into a time-independent part $H_0$, given explicitly in Eq.~(\ref{eq:h0}), and a time-periodic contribution whose Fourier components read
\begin{align}
    V^{(n)} = \sum_{i \neq j} \frac{J_{i,j}}{2\pi n} \Big[ \,
    & \sin\!\big(n\omega(\tau_1+\tau_2)\big)\big(\sigma_i^x \sigma_j^x - \sigma_i^y \sigma_j^y\big)
        \nonumber\\
        &+ \sin\!\big(n\omega\tau_1\big)\big(\sigma_i^y \sigma_j^y - \sigma_i^z \sigma_j^z\big)
    \,\Big],
\end{align}
with $V^{(n)\dagger} = V^{(n)}$ and $V^{(0)} \equiv 0$.
Within the high-frequency Floquet formalism, the effective Hamiltonian governing the stroboscopic dynamics over one driving period is given by
\begin{align}
    H_{\text{eff}} &= H_0 
    + \frac{1}{\omega} \sum_{n=1}^\infty \frac{1}{n} [V^{(n)}, V^{(n)\dagger}] \nonumber\\
    &+ \frac{1}{2\omega^2} \sum_{n=1}^\infty \frac{1}{n^2}  [[V^{(n)}, H_0], V^{(n)\dagger}]  \nonumber\\
    &+ \frac{1}{\omega^2} \sum_{n,m=1}^\infty \frac{1}{n(n+m)}  
    [V^{(n)}, [V^{(m)}, V^{(n+m)\dagger}]]  \nonumber\\
    &+ \text{h.c.} + \mathcal{O}(1/\omega^3).
\label{eq:effective_uptosecond}
\end{align}

In the present case, since $V^{(n)} = V^{(n)\dagger}$, the first-order correction in $1/\omega$ vanishes identically. 
As a result, the leading nontrivial corrections to $H_0$ arise at second order in the inverse driving frequency and scale as $1/\omega^2$.
This expansion follows the standard high-frequency Floquet formalism described in Ref.~\cite{goldman_periodically_2014}.

\section{Second order coefficients}
\label{appendix_coefficients}
As shown in Appendix~\ref{appendix_floquet}, the leading nonvanishing corrections to the effective Hamiltonian arise at second order in the high-frequency expansion and scale as $1/\omega^2$. 
In this appendix we provide the explicit expressions of the corresponding second-order effective Hamiltonian, which consists of both two-spin and four-spin interaction terms ($H^{(2)}_{\text{eff}}/\omega^2 = H^{2\sigma} + H^{4\sigma}$), given in Eqs.~(\ref{eq:twospins_ham}) and~(\ref{eq:fourspin_ham}). 
The coefficients appearing in the two-spin second-order Hamiltonian $H^{2\sigma}$ are given by
\begin{equation}
\begin{split}
    \mathcal{J}^x_{j,k} &= \frac{8}{\omega^2\pi^2}
    \sum_{i\neq j,k}\sum_{\substack{n=1 \\ m = 0}}^{\infty}
    \frac{ J_{i,j}\!\left(\Lambda^{(n,m)}_{kji}A^{(n)}_z-\Lambda^{(n,m)}_{kij}A^{(n)}_y \right)}{n^2(n+m)^2},\\
    \mathcal{J}^y_{j,k} &= \frac{8}{\omega^2\pi^2}
    \sum_{i\neq j,k}\sum_{\substack{n=1 \\ m = 0}}^{\infty}
    \frac{ J_{i,j}\!\left( \Lambda^{(n,m)}_{ikj}A^{(n)}_x - \Lambda^{(n,m)}_{jki}A^{(n)}_z \right)}{n^2(n+m)^2},\\
    \mathcal{J}^z_{j,k} &= \frac{8}{\omega^2\pi^2}
    \sum_{i\neq j,k}\sum_{\substack{n=1 \\ m = 0}}^{\infty}
    \frac{ J_{i,j}\!\left( \Lambda^{(n,m)}_{jik}A^{(n)}_y  -\Lambda^{(n,m)}_{ijk}A^{(n)}_x\right)}{n^2(n+m)^2}.
\end{split}
\end{equation}
Here $J_{i,j}$ denotes the native dipolar couplings introduced in Eq.~(\ref{eq:XY}). 
The remaining coefficients entering the above expressions are defined as
\begin{equation}
\label{eq:gen:substitutions}
\begin{split}
    A_x^{(n)} &= \sin \!\big[n\omega(\tau_1 + \tau_2)\big],\\
    A_y^{(n)} &= \sin (n\omega\tau_1)-\sin \!\big[n\omega(\tau_1 + \tau_2)\big],\\
    A_z^{(n)} &= -\sin (n\omega\tau_1).
\end{split}
\end{equation}

The coefficients $\Lambda^{(n,m)}_{ijk}$ take different forms depending on whether $m=0$ or $m\neq 0$. 
For $m=0$ they read
\begin{equation}
\begin{split}
        \Lambda^{(n,0)}_{ijk} =\;&
        \Big[A^{(n)}_z(\tau_1+\tau_2)-A^{(n)}_x(\tau_2+\tau_3)\Big]\frac{J_{i,j}J_{j,k}}{T} \\
        + &\Big[A^{(n)}_x(\tau_1+\tau_3) -A^{(n)}_y(\tau_1+\tau_2)\Big]\frac{J_{i,k}J_{k,j}}{T}\\
        + &\Big[A^{(n)}_y(\tau_2+\tau_3) -A^{(n)}_z(\tau_1+\tau_3)\Big]\frac{J_{j,i}J_{i,k}}{T},
\end{split}
\end{equation}
whereas for $m\neq 0$ one finds
\begin{equation}
\begin{split}
\label{eq:lambda_mneq0}
        \Lambda^{(n,m)}_{ijk} =\;&
        \Big[A^{(n+m)}_z A^{(m)}_x-A^{(n+m)}_x A^{(m)}_z\Big]\frac{J_{i,j}J_{j,k}}{\pi\, m} \\
        + &\Big[A^{(n+m)}_x A^{(m)}_y -A^{(n+m)}_y A^{(m)}_x\Big]\frac{J_{i,k}J_{k,j}}{\pi\, m} \\
        + &\Big[A^{(n+m)}_y A^{(m)}_z -A^{(n+m)}_z A^{(m)}_y \Big]\frac{J_{j,i}J_{i,k}}{\pi\, m}.
\end{split}
\end{equation}

The four-spin contribution to the second-order effective Hamiltonian can be originally written as
\begin{align}
    H^{4\sigma} = \sum_{h\neq i\neq j\neq k} \Big(
    &c^{yz}_{hijk}\,\sigma^y_h\sigma^y_i\sigma^z_j\sigma^z_k 
    + c^{zx}_{hijk}\,\sigma^z_h\sigma^z_i\sigma^x_j\sigma^x_k \nonumber\\
    &+ c^{xy}_{hijk}\,\sigma^x_h\sigma^x_i\sigma^y_j\sigma^y_k 
    \Big),
\end{align}
where the coefficients are given by
\begin{equation}
\begin{split}
    c^{yz}_{hijk} &= \frac{8}{\omega^2\pi^2}
    \sum_{\substack{n=1\\ m=0}}^{\infty}
    \frac{J_{h,j}\!\left(\Lambda^{(n,m)}_{hik}A_z^{(n)}-\Lambda^{(n,m)}_{jik}A_y^{(n)}\right)}{n^2( n+m)^2},  \\
    c^{zx}_{hijk} &= \frac{8}{\omega^2\pi^2}
    \sum_{\substack{n=1\\ m=0}}^{\infty}
    \frac{J_{h,j}\!\left(\Lambda^{(n,m)}_{khi}A_x^{(n)}-\Lambda^{(n,m)}_{kji}A_z^{(n)}\right)}{n^2( n+m)^2},  \\
    c^{xy}_{hijk} &= \frac{8}{\omega^2\pi^2}
    \sum_{\substack{n=1\\ m=0}}^{\infty}
    \frac{J_{h,j}\!\left(\Lambda^{(n,m)}_{ikh}A_y^{(n)}-\Lambda^{(n,m)}_{ikj}A_x^{(n)}\right)}{n^2( n+m)^2}.
\end{split}
\end{equation}

The spin operators appearing in $H^{4\sigma}$ are symmetric under the exchanges $h\leftrightarrow i$ and $j\leftrightarrow k$. 
It is therefore convenient to introduce symmetrized coefficients
\begin{equation}
    C_{hi;jk} = c_{hijk} + c_{hikj} + c_{ihjk} + c_{ihkj},
\end{equation}
in terms of which the four-spin Hamiltonian takes the compact form reported in Eq.~\eqref{eq:fourspin_ham}. 
The restricted summation in that equation, together with the condition $\{h,i\}\cap\{j,k\}=\varnothing$, accounts for the fact that the coefficients are not symmetric under the exchange of the two index pairs $(h,i)\leftrightarrow(j,k)$.

\begin{figure}
    \centering
    \includegraphics[width=\linewidth]{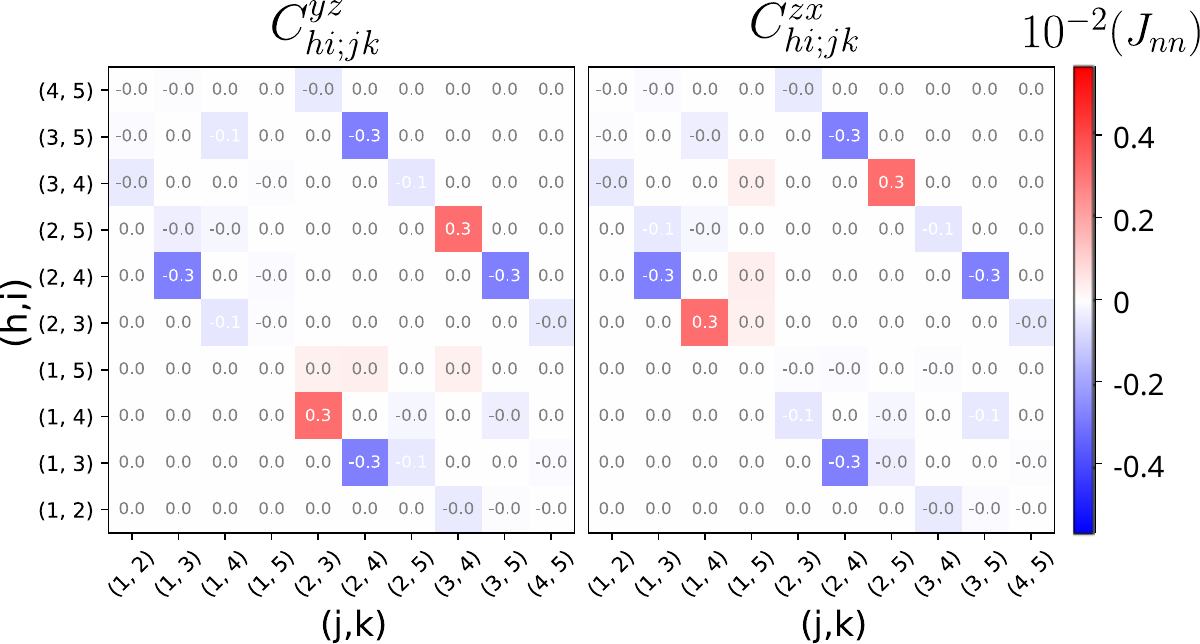}
    \caption{
    \justifying
    Visualization of the four-body interaction coefficients $C^{yz}_{hi;jk}$ and $C^{zx}_{hi:jk}$ appearing in $H^{4\sigma}$ (Eq.~\ref{eq:fourspin_ham}) for a 5-site chain with dipolar couplings ($J_{ij} \propto 1/|i-j|^3$), evaluated at $\Gamma = 0.01$. The driving protocol is characterized by WAHUHA parameters $\tau_1 = 0$, $\tau_2/T = 0.33$, and $\tau_3/T = 0.17$. The $y$-axis of each subplot labels all index pairs $(h,i)$ and the $x$-axis all pairs $(j,k)$. 
    The equivalence between $C^{yz}_{hi;jk}$ and $C^{zx}_{jk;hi}$ across the plots illustrates that the symmetry condition for the four-spin coefficients is satisfied when $\tau_1 \to 0$. As discussed in Appendix~\ref{appendix_coefficients}, the full pseudo-$SO(2)$ symmetry requires in addition $\tau_2 = \tau_3$, which enforces an $SO(2)$ symmetry already at zeroth order.}
    \label{fig:fourspincoeffs}
\end{figure}

The emergence of a pseudo-$SO(2)$ symmetry at second order requires the conditions $\mathcal{J}^x_{j,k}=\mathcal{J}^y_{j,k}$ for all $j,k$ and $C^{yz}_{hi;jk}=C^{zx}_{jk;hi}$ for all $h,i,j,k$. 
These constraints imply $A_z^{(n)}=0$ and $A_y^{(n)}=-A_x^{(n)}$, which can only be satisfied in the limit $\tau_1\rightarrow 0$ (see Fig.~\ref{fig:fourspincoeffs}). 
This choice must be combined with the condition $\tau_2=\tau_3$, which enforces an $SO(2)$ symmetry already at zeroth order.

With this choice of parameters, the definitions of the coefficients simplify dramatically: one can verify that all quantities $\Lambda^{(n,m)}$ with $m\neq 0$, defined in Eq.~(\ref{eq:lambda_mneq0}), vanish. The remaining sums over $n$ (with $m=0$) reduce to
\begin{equation*} 
\frac{8}{(\omega\pi)^2}\sum_{n=1}^{\infty}\frac{(A^{(n)}_x)^2}{n^4} = \frac{8}{(\omega\pi)^2}\sum_{n=1}^{\infty}\frac{\sin^2(n\pi/2)}{n^4} = \frac{1}{12}\left(\frac{\pi}{\omega}\right)^2, 
\end{equation*}
up to products of the couplings $J$. Introducing the parameter $\Gamma\equiv \frac{1}{6}\left(\frac{J_{\text{nn}} \pi}{\omega}\right)^2$, the Hamiltonian~(\ref{eq:effective_uptosecond}) can be cast in the form $H_{\text{eff}}=\sum_{jk}H_{j>k} + \mathcal{O}(1/\omega^3)$, with:
\begin{widetext}
\begin{align}
{H_{j>k}} =\;&
{J_{j,k}}\left(\sigma^+_j\sigma^-_k + \sigma^-_j\sigma^+_k + \sigma^z_j\sigma^z_k\right)
\nonumber\\[0.3em]
&-\frac{\Gamma}{2J_{\mathrm{nn}}^2}\sum_{i\neq j,k}
\Big(
J^2_{i,j}J_{j,k} + J^2_{i,k}J_{j,k}
- 2J_{i,j}J_{i,k}J_{j,k}
- J^2_{i,j}J_{i,k} - J^2_{i,k}J_{i,j}
\Big)
\left(\sigma^+_j\sigma^-_k + \sigma^-_j\sigma^+_k\right)
\nonumber\\[0.3em]
&-\frac{\Gamma}{2J_{\mathrm{nn}}^2}\sum_{i\neq j,k}
\Big(
J^2_{i,j}J_{j,k} + J^2_{i,j}J_{i,k}
+ J^2_{i,k}J_{j,k} + J^2_{i,k}J_{i,j}
- 2J_{i,j}J_{i,k}J_{j,k}
\Big)
\:\sigma^z_j\sigma^z_k
\nonumber\\[0.3em]
&-\frac{\Gamma}{J_{\mathrm{nn}}^2} \sum_{h>i}
\Bigg[
 J_{h,j}J_{i,k}J_{h,k}
+J_{h,k}J_{i,j}J_{h,j}
+J_{i,j}J_{h,k}J_{i,k}
+J_{i,k}J_{h,j}J_{i,j}
\nonumber\\
&\qquad\qquad
+2\!\left( 
 J_{h,i}J_{h,j}J_{h,k}
+J_{i,j}J_{h,i}J_{i,k} 
-J_{h,j}J_{h,i}J_{i,k} 
-J_{h,k}J_{h,i}J_{i,j}\right)\Bigg]
\sigma^z_h\sigma^z_i
\left(\sigma^+_j\sigma^-_k + \sigma^-_j\sigma^+_k\right)
\nonumber\\[0.3em]
&+\frac{\Gamma}{2J_{\mathrm{nn}}^2}\sum_{h>i}
\Bigg[
+J_{h,j}J_{i,k}J_{h,k}
+J_{h,k}J_{i,j}J_{h,j}
+J_{i,j}J_{h,k}J_{i,k}
+J_{i,k}J_{h,j}J_{i,j}
\nonumber\\
&\qquad\quad\quad\:\:
+
J_{h,j}J_{h,i}J_{i,k}
+J_{h,k}J_{h,i}J_{i,j}
+J_{h,j}J_{i,k}J_{j,k}
+J_{h,k}J_{i,j}J_{j,k}
\nonumber\\
&\qquad\quad\quad\:\:
-\Big(
J_{h,i}J_{h,j}J_{h,k}
+J_{i,j}J_{h,i}J_{i,k}
+J_{h,j}J_{i,j}J_{j,k}
+J_{h,k}J_{i,k}J_{j,k}
\Big)
\Bigg]
\sigma^x_h\sigma^x_i\sigma^y_j\sigma^y_k .
\label{eq:general_hamiltonian}
\end{align}
\end{widetext}

This expression remains valid for arbitrary lattice geometries and interaction profiles $J_{i,j}$.
The one-dimensional nearest-neighbor limit relevant for Eqs.~\eqref{eq:effective_1d_nn} and~\eqref{eq:symm-break-terms} is discussed explicitly below.

A convenient alternative set of parameters is obtained by taking $\tau_3\to 0$ while imposing $\tau_1=\tau_2$ (that is, $A_x^{(n)}=0$ and $A_z^{(n)}=-A_y^{(n)}$). In this case the two central pulses with phases $\phi=-\pi/2$ and $\phi=+\pi/2$ become coincident at mid period and cancel each other, so that the protocol again reduces to a two-pulse sequence per period, now consisting of $\phi=0$ and $\phi=\pi$ applied at $t=T/4$ and $t=3T/4$. Crucially, this implementation avoids the need for a pulse at $t=0$, leaving a time window of $T/4$ after the quench before the first pulse is applied. The resulting effective Hamiltonian is equivalent to the one discussed above up to a global spin rotation (i.e., a relabeling of spin axes); in this paper we nonetheless adopt the convention in which the pseudo-$SO(2)$ symmetry lies in the $x$--$y$ plane, which is the most common choice in the literature.

\subsection{Nearest-neighbor reduction in one dimension}
\label{app:nn_reduction}

We now specialize Eq.~\eqref{eq:general_hamiltonian} to a one-dimensional chain with nearest-neighbor couplings, $J_{ij}=J_{\rm {nn}}$ for $|i-j|=1$ and $J_{ij}=0$ otherwise. 
Using $J_{\rm {nn}}$ as the energy unit, we summarize how the general expression reduces to Eqs.~\eqref{eq:effective_1d_nn} and~\eqref{eq:symm-break-terms}.

The reduction is most transparent by observing that every second-order coefficient in Eq.~\eqref{eq:general_hamiltonian} is a product of three native couplings. 
Therefore, in the nearest-neighbor limit, a contribution survives only if all three factors correspond to nearest-neighbor bonds. 
This imposes strong geometric constraints on the relative positions of the indices and eliminates all terms requiring one site to be simultaneously connected to three distinct sites.
For the terms involving two spin operators, this selection rule is illustrated schematically in Fig.~\ref{fig:nn_reduction_rule}.


\begin{figure}[t]
\centering
\begin{tikzpicture}[scale=0.85]
\tikzstyle{site}=[circle, fill=black, inner sep=1.4pt]
\tikzstyle{bond}=[gray, thick]
\tikzstyle{longbond}=[gray, thick, dashed]

\draw[bond] (0,0) -- (1,0);
\draw[bond] (1,0) -- (2,0);
\node[site,label=below:{\small $i$}] at (0,0) {};
\node[site,label=below:{\small $j$}] at (1,0) {};
\node[site,label=below:{\small $k$}] at (2,0) {};
\draw[gray, thick, double] (1,0.12) -- (2,0.12);
\node[align=center] at (1,-1) {\small $|j-k|=1$\\[-0.1em]\small n.n. renorm.};

\draw[bond] (3,0) -- (4,0);
\draw[bond] (4,0) -- (5,0);
\node[site,label=below:{\small $j$}] at (3,0) {};
\node[site,label=below:{\small $i$}] at (4,0) {};
\node[site,label=below:{\small $k$}] at (5,0) {};
\draw[gray, thick, double] (3,0.35) .. controls (4,0.75) .. (5,0.35);
\node[align=center] at (4,-1) {\small $|j-k|=2$\\[-0.1em]\small generated};

\draw[bond] (6,0) -- (7,0);
\draw[bond] (7,0) -- (8,0);
\draw[bond] (8,0) -- (9,0);
\node[site,label=below:{\small $j$}] at (6,0) {};
\node[site,label=below:{\small $i$}] at (7,0) {};
\node[site] at (8,0) {};
\node[site,label=below:{\small $k$}] at (9,0) {};
\draw[longbond] (6,0.35) .. controls (7.5,1.0) .. (9,0.35);
\node[align=center] at (7.5,-1) {\small long bond\\[-0.1em]\small vanishes};
\end{tikzpicture}
\caption{
\justifying
Schematic selection rule for the nearest-neighbor reduction of the terms involving two spin operators in Eq.~\eqref{eq:general_hamiltonian}. 
Double lines denote the effective coupling generated between the external sites, while single lines denote the native couplings. 
Only products made entirely of nearest-neighbor bonds survive; hence these terms either renormalize nearest-neighbor couplings or generate next-nearest-neighbor ones. 
The four-spin-operator terms follow the same selection principle but have different allowed topologies.
}
\label{fig:nn_reduction_rule}
\end{figure}


In this part of the Hamiltonian, the only nonvanishing contributions occur for $|j-k|=1$ and $|j-k|=2$: the former renormalize the leading-order XXZ Hamiltonian, while the latter generate the second line of Eq.~\eqref{eq:effective_1d_nn}. Explicitly, 

\begin{equation}
\label{eq:app_twospin_nn}
\begin{split}
H^{2\sigma}_{\rm{nn}}
&=
-\Gamma\sum_i
\left(
\sigma_i^+\sigma_{i+1}^-
+\sigma_i^-\sigma_{i+1}^+
+\sigma_i^z\sigma_{i+1}^z
\right)\\
&+
\Gamma\sum_i
\left(
\sigma_i^+\sigma_{i+2}^-
+\sigma_i^-\sigma_{i+2}^+
-\sigma_i^z\sigma_{i+2}^z
\right).
\end{split}
\end{equation}
Adding the zeroth-order contribution gives the first two lines of Eq.~\eqref{eq:effective_1d_nn}.

We next consider the four-spin terms proportional to
$\sigma_h^z\sigma_i^z(\sigma_j^+\sigma_k^-+\mathrm{h.c.})$. 
In the nearest-neighbor chain, the allowed index configurations can be classified according to the distance between the hopping sites $j$ and $k$. 
First-neighbor correlated hopping is absent because it would require at least one vanishing native coupling. 
For $|j-k|=2$, the surviving configurations are those where the two $z$-operators occupy sites adjacent to the hopping pair, giving
\begin{equation}
\label{eq:app_fourspin_nn_2}
-\Gamma\sum_i
\left(
\sigma^z_{i-1}\sigma^z_{i+1}
+
\sigma^z_{i+1}\sigma^z_{i+3}
\right)
\left(
\sigma_i^+\sigma_{i+2}^-
+
\sigma_i^-\sigma_{i+2}^+
\right).
\end{equation}
For $|j-k|=3$, the two intermediate sites carry the $z$-operators, giving
\begin{equation}
\label{eq:app_fourspin_nn_3}
2\Gamma\sum_i
\sigma^z_{i+1}\sigma^z_{i+2}
\left(
\sigma_i^+\sigma_{i+3}^-
+
\sigma_i^-\sigma_{i+3}^+
\right).
\end{equation}
Equations~\eqref{eq:app_fourspin_nn_2} and~\eqref{eq:app_fourspin_nn_3} correspond respectively to the third and fourth lines of Eq.~\eqref{eq:effective_1d_nn}.

Finally, the $xxyy$ part of Eq.~\eqref{eq:general_hamiltonian} survives only when the four indices occupy four consecutive sites. 
After summing over the inequivalent permutations and rewriting the result in terms of $\sigma^\pm$, one obtains
\begin{align}
\label{eq:app_xxyy_nn}
H^{xxyy}_{\rm{nn}}
&=
2\Gamma\sum_i
\left(
\sigma_i^+\sigma_{i+1}^+
\sigma_{i+2}^-\sigma_{i+3}^-
+
\sigma_i^-\sigma_{i+1}^-
\sigma_{i+2}^+\sigma_{i+3}^+
\right)
\nonumber\\
&-2\Gamma\sum_i
\left(
\sigma_i^+\sigma_{i+1}^+
\sigma_{i+2}^+\sigma_{i+3}^+
+
\sigma_i^-\sigma_{i+1}^-
\sigma_{i+2}^-\sigma_{i+3}^-
\right).
\end{align}
The first line conserves the total magnetization and gives the last line of Eq.~\eqref{eq:effective_1d_nn}. 
The second line changes $N_\uparrow$ by four units and is the only term that breaks the emergent $SO(2)$ symmetry in the one-dimensional nearest-neighbor limit, yielding Eq.~\eqref{eq:symm-break-terms}. 
All other terms in Eq.~\eqref{eq:general_hamiltonian} vanish in this limit because at least one of the native couplings appearing in the corresponding product is zero.

\section{Floquet theory benchmark}
\label{appendix: benchmark}

We validate the second-order effective Floquet Hamiltonian \eqref{eq:effective_1d_nn} with closed boundary conditions by comparing its dynamics with the exact one obtained via the time-dependent evolution under the full Hamiltonian \eqref{eq: full-ham}. 
This comparison focuses on the magnetization dynamics starting from a trimer state, i.e. a state in which three neighboring spins are flipped on top of a fully polarized background.

\begin{figure}
    \centering
    \includegraphics[width=\linewidth]{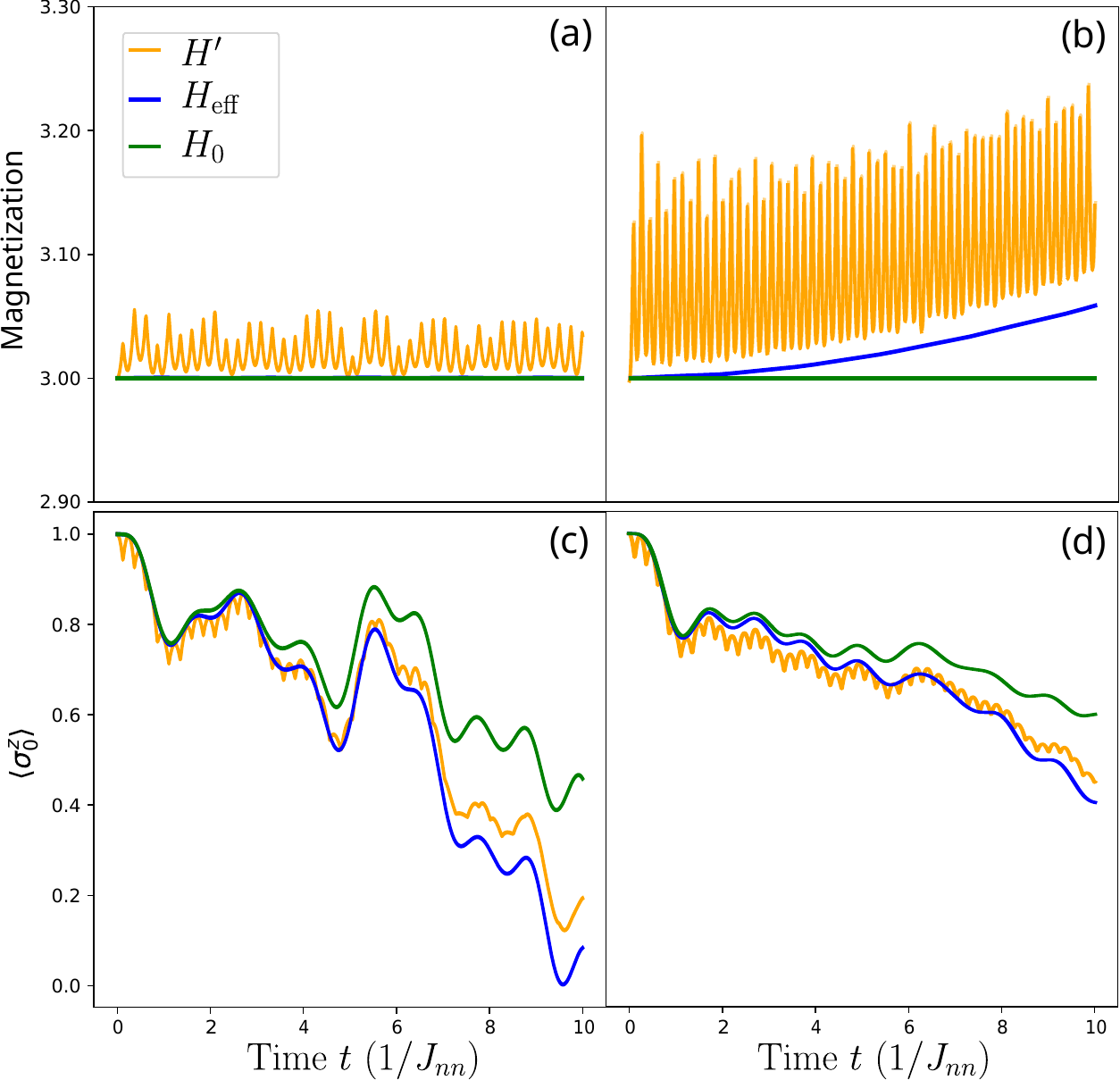}
    \caption{
        \justifying
        Time evolution of global (top panels) and local magnetization (bottom panels) in finite spin chains with $L=7$ (left panels) and $L=15$ (right panels), starting from a trimer initial state localized at the center of the chain and corresponding to $\langle N_\uparrow \rangle = 3$. Closed boundary conditions are used in all simulations. In all panels, we compare the dynamics generated by the exact driven Hamiltonian (orange), the second-order effective Hamiltonian (Eqs.~\eqref{eq:effective_1d_nn}-\eqref{eq:symm-break-terms} in the main text) (blue), and the leading-order XXZ model (green line). The driving parameters are chosen such that $\Gamma=0.01$ for the $L=7$ panels \textbf{(a)}-\textbf{(c)} and $\Gamma=0.005$ for the $L=15$ panels \textbf{(b)}-\textbf{(d)}.
    }
    \label{fig:benchmark}
\end{figure}

For relatively small system sizes (up to 7 spins) the two evolutions show excellent agreement even at moderate driving frequencies $\omega/J_{\text{nn}}\sim 6$. 
In this regime, the parameter $\Gamma$ reaches values as high as $0.02$ without compromising the accuracy of the effective description \eqref{eq:effective_1d_nn} (see Fig.~\ref{fig:benchmark}(a)).

For larger systems, higher-order terms can become more visible, since global observables accumulate small corrections over many sites and over longer timescales.
As a consequence, the range of parameters where the second-order effective Hamiltonian provides a uniformly quantitative description becomes more restricted as the system size increases.

For example, when studying the behavior of total magnetization, the effective Hamiltonian (\ref{eq:effective_1d_nn}) already includes the leading symmetry-breaking channel at order $1/\omega^2$.
This already provides a substantial improvement over the leading-order XXZ Hamiltonian, which strictly conserves magnetization and therefore cannot capture this drift at all. 
At the same time, in longer chains this leading symmetry-breaking contribution does not fully account for the magnetization variation observed in the exact driven dynamics.
This indicates that the second-order effective Hamiltonian should not be regarded as a uniformly quantitative description of all observables in this regime, especially for global quantities that are sensitive to small magnetization-changing processes accumulated over the whole system. 
As expected, reducing $\Gamma$, and therefore increasing the driving frequency, improves the agreement with the exact driven dynamics; see Fig.~\ref{fig:benchmark}(b).

On the other hand, the discrepancies observed in the total magnetization do not affect all observables in the same way. 
Local quantities, which are the main diagnostics used in the main text to identify the propagation of the trimer excitation, are captured more accurately by the second-order effective model. 
This is illustrated in Fig.~\ref{fig:benchmark}(d), where the local magnetization at the center of the chain for $L=15$ follows the exact driven dynamics substantially better than the leading-order XXZ Hamiltonian. 
We therefore use the second-order effective Hamiltonian primarily as a controlled guide to identify the relevant Floquet-induced processes and the associated propagation scales, while the visibility of the trimer signal is assessed directly from the driven dynamics in the main text.

\section{Numerical methods}
\label{appendix:numerics}

All numerical simulations presented in this work are based on exact diagonalization and exact unitary time evolution of finite spin chains. 
The dynamics is initialized from product states corresponding to localized excitations, typically trimers placed at the center of chain with closed boundary conditions.

The construction of the spin Hamiltonians and the time evolutions were implemented using the QuSpin library~\cite{weinberg2017quspin, weinberg2019quspin}, together with standard numerical routines from Python libraries. 
Momentum-resolved spectra were obtained via exact diagonalization within symmetry-resolved Hilbert-space sectors.

\newpage

\bibliographystyle{apsrev4-1}
\bibliography{biblio}

@article{lu2026probing,
  title={Probing Coherent Many-Body Spin Dynamics in a Molecular Tweezer Array Quantum Simulator},
  author={Lu, Yukai and Holland, Connor M and Welsh, Callum L and Chen, Xing-Yan and Cheuk, Lawrence W},
  journal={arXiv:2603.19090},
  year={2026},
  url={https://arxiv.org/abs/2603.19090}
}

@article{kuang2026dynamics,
  title={Dynamics of antiferromagnetic Dimers in Rydberg Atom Chains},
  author={Kuang, Feng-Yuan and Li, Lin and Li, Weibin},
  journal={arXiv preprint arXiv:2601.15866},
  year={2026}
}

@article{Zhou2024,
  title = {Robust Hamiltonian Engineering for Interacting Qudit Systems},
  author = {Zhou, Hengyun and Gao, Haoyang and Leitao, Nathaniel T. and Makarova, Oksana and Cong, Iris and Douglas, Alexander M. and Martin, Leigh S. and Lukin, Mikhail D.},
  journal = {Phys. Rev. X},
  volume = {14},
  issue = {3},
  pages = {031017},
  numpages = {33},
  year = {2024},
  month = {Jul},
  publisher = {American Physical Society},
  doi = {10.1103/PhysRevX.14.031017},
  url = {https://link.aps.org/doi/10.1103/PhysRevX.14.031017}
}

@article{Hazzard2024,
	author = {Cornish, Simon L. and Tarbutt, Michael R. and Hazzard, Kaden R. A.},
	date = {2024/05/01},
	doi = {10.1038/s41567-024-02453-9},
	id = {Cornish2024},
	isbn = {1745-2481},
	journal = {Nature Physics},
	number = {5},
	pages = {730--740},
	title = {Quantum computation and quantum simulation with ultracold molecules},
	url = {https://doi.org/10.1038/s41567-024-02453-9},
	volume = {20},
	year = {2024}
}

@article{Zoller2006,
	author = {Micheli, A. and Brennen, G. K. and Zoller, P.},
	date = {2006/05/01},
	doi = {10.1038/nphys287},
	id = {Micheli2006},
	isbn = {1745-2481},
	journal = {Nature Physics},
	number = {5},
	pages = {341--347},
	title = {A toolbox for lattice-spin models with polar molecules},
	url = {https://doi.org/10.1038/nphys287},
	volume = {2},
	year = {2006}
}

@article{chomaz2022dipolar,
  title={Dipolar physics: a review of experiments with magnetic quantum gases},
  author={Chomaz, Lauriane and Ferrier-Barbut, Igor and Ferlaino, Francesca and Laburthe-Tolra, Bruno and Lev, Benjamin L and Pfau, Tilman},
  journal={Reports on Progress in Physics},
  volume={86},
  number={2},
  pages={026401},
  year={2022},
  publisher={IOP Publishing},
  url={https://iopscience.iop.org/article/10.1088/1361-6633/aca814/meta}
}

@article{Ye2024,
	author = {Miller, Calder and Carroll, Annette N. and Lin, Junyu and Hirzler, Henrik and Gao, Haoyang and Zhou, Hengyun and Lukin, Mikhail D. and Ye, Jun},
	date = {2024/09/01},
	doi = {10.1038/s41586-024-07883-2},
	id = {Miller2024},
	isbn = {1476-4687},
	journal = {Nature},
	number = {8029},
	pages = {332--337},
	title = {Two-axis twisting using Floquet-engineered XYZ spin models with polar molecules},
	url = {https://doi.org/10.1038/s41586-024-07883-2},
	volume = {633},
	year = {2024}
}

@article{Pavesic2025,
  title = {Constrained dynamics and confinement in the two-dimensional quantum Ising model},
  author = {Pave\ifmmode \check{s}\else \v{s}\fi{}i\ifmmode \acute{c}\else \'{c}\fi{}, Luka and Jaschke, Daniel and Montangero, Simone},
  journal = {Phys. Rev. B},
  volume = {111},
  issue = {14},
  pages = {L140305},
  numpages = {7},
  year = {2025},
  month = {Apr},
  publisher = {American Physical Society},
  doi = {10.1103/PhysRevB.111.L140305},
  url = {https://link.aps.org/doi/10.1103/PhysRevB.111.L140305}
}

@article{Cuadra2025,
	author = {Gonz{\'a}lez-Cuadra, Daniel and Hamdan, Majd and Zache, Torsten V. and Braverman, Boris and Kornja{\v c}a, Milan and Lukin, Alexander and Cant{\'u}, Sergio H. and Liu, Fangli and Wang, Sheng-Tao and Keesling, Alexander and Lukin, Mikhail D. and Zoller, Peter and Bylinskii, Alexei},
	date = {2025/06/01},
	doi = {10.1038/s41586-025-09051-6},
	id = {Gonz{\'a}lez-Cuadra2025},
	isbn = {1476-4687},
	journal = {Nature},
	number = {8067},
	pages = {321--326},
	title = {Observation of string breaking on a (2 + 1)D Rydberg quantum simulator},
	url = {https://doi.org/10.1038/s41586-025-09051-6},
	volume = {642},
	year = {2025}
}

@article{Surace2020,
  title = {Lattice Gauge Theories and String Dynamics in Rydberg Atom Quantum Simulators},
  author = {Surace, Federica M. and Mazza, Paolo P. and Giudici, Giuliano and Lerose, Alessio and Gambassi, Andrea and Dalmonte, Marcello},
  journal = {Phys. Rev. X},
  volume = {10},
  issue = {2},
  pages = {021041},
  numpages = {14},
  year = {2020},
  month = {May},
  publisher = {American Physical Society},
  doi = {10.1103/PhysRevX.10.021041},
  url = {https://link.aps.org/doi/10.1103/PhysRevX.10.021041}
}

@article{calabrese2017,
	author = {Kormos, Marton and Collura, Mario and Tak{\'a}cs, Gabor and Calabrese, Pasquale},
	date = {2017/03/01},
	doi = {10.1038/nphys3934},
	id = {Kormos2017},
	isbn = {1745-2481},
	journal = {Nature Physics},
	number = {3},
	pages = {246--249},
	title = {Real-time confinement following a quantum quench to a non-integrable model},
	url = {https://doi.org/10.1038/nphys3934},
	volume = {13},
	year = {2017}
}

@article{Petrosyan2018,
  title = {Mobile bound states of Rydberg excitations in a lattice},
  author = {Letscher, Fabian and Petrosyan, David},
  journal = {Phys. Rev. A},
  volume = {97},
  issue = {4},
  pages = {043415},
  numpages = {13},
  year = {2018},
  month = {Apr},
  publisher = {American Physical Society},
  doi = {10.1103/PhysRevA.97.043415},
  url = {https://link.aps.org/doi/10.1103/PhysRevA.97.043415}
}

@article{qiao2025kinetically,
  title={Kinetically-induced bound states in a frustrated Rydberg tweezer array},
  author={Qiao, Mu and Martin, Romain and Homeier, Lukas and Morera, Ivan and G{\'e}ly, Bastien and Klein, Lukas and Chew, Yuki Torii and Barredo, Daniel and Lahaye, Thierry and Demler, Eugene and others},
  journal={arXiv preprint arXiv:2510.17183},
  year={2025},
  url={https://arxiv.org/abs/2510.17183}
}

@article{Defenu2023,
	author = {Defenu, Nicol\`o and Donner, Tobias and Macr\`{\i}, Tommaso and Pagano, Guido and Ruffo, Stefano and Trombettoni, Andrea},
	doi = {10.1103/RevModPhys.95.035002},
	issue = {3},
	journal = {Rev. Mod. Phys.},
	month = {Aug},
	numpages = {70},
	pages = {035002},
	publisher = {American Physical Society},
	title = {Long-range interacting quantum systems},
	url = {https://link.aps.org/doi/10.1103/RevModPhys.95.035002},
	volume = {95},
	year = {2023}
}

@article{Keesling2019,
	author = {Keesling, Alexander and Omran, Ahmed and Levine, Harry and Bernien, Hannes and Pichler, Hannes and Choi, Soonwon and Samajdar, Rhine and Schwartz, Sylvain and Silvi, Pietro and Sachdev, Subir and Zoller, Peter and Endres, Manuel and Greiner, Markus and Vuleti{\ifmmode\acute{c}\else\'{c}\fi}, Vladan and Lukin, Mikhail D.},
	title = {{Quantum Kibble{\textendash}Zurek mechanism and critical dynamics on a programmable Rydberg simulator}},
	journal = {Nature},
	volume = {568},
	pages = {207--211},
	year = {2019},
	month = apr,
	issn = {1476-4687},
	publisher = {Nature Publishing Group},
	doi = {10.1038/s41586-019-1070-1}
}

@article{Tecer2024,
	author = {Te\ifmmode \check{c}\else \v{c}\fi{}er, Matija and Di Liberto, Marco and Silvi, Pietro and Montangero, Simone and Romanato, Filippo and Calaj\'o, Giuseppe},
	doi = {10.1103/PhysRevLett.132.163602},
	issue = {16},
	journal = {Phys. Rev. Lett.},
	month = {Apr},
	numpages = {7},
	pages = {163602},
	publisher = {American Physical Society},
	title = {Strongly Interacting Photons in 2D Waveguide QED},
	url = {https://link.aps.org/doi/10.1103/PhysRevLett.132.163602},
	volume = {132},
	year = {2024}
}

@article{Struck2012,
	author = {Struck, J. and \"Olschl\"ager, C. and Weinberg, M. and Hauke, P. and Simonet, J. and Eckardt, A. and Lewenstein, M. and Sengstock, K. and Windpassinger, P.},
	doi = {10.1103/PhysRevLett.108.225304},
	issue = {22},
	journal = {Phys. Rev. Lett.},
	month = {May},
	numpages = {5},
	pages = {225304},
	publisher = {American Physical Society},
	title = {Tunable Gauge Potential for Neutral and Spinless Particles in Driven Optical Lattices},
	url = {https://link.aps.org/doi/10.1103/PhysRevLett.108.225304},
	volume = {108},
	year = {2012}}

@article{Sun2023,
	author = {Sun, Bo-Ye and Goldman, Nathan and Aidelsburger, Monika and Bukov, Marin},
	doi = {10.1103/PRXQuantum.4.020329},
	issue = {2},
	journal = {PRX Quantum},
	month = {May},
	numpages = {24},
	pages = {020329},
	publisher = {American Physical Society},
	title = {Engineering and Probing Non-Abelian Chiral Spin Liquids Using Periodically Driven Ultracold Atoms},
	url = {https://link.aps.org/doi/10.1103/PRXQuantum.4.020329},
	volume = {4},
	year = {2023}}

@article{Zhao2023,
	author = {Zhao, Luheng and Lee, Michael Dao Kang and Aliyu, Mohammad Mujahid and Loh, Huanqian},
	date = {2023/11/06},
	doi = {10.1038/s41467-023-42899-8},
	id = {Zhao2023},
	isbn = {2041-1723},
	journal = {Nature Communications},
	number = {1},
	pages = {7128},
	title = {Floquet-tailored Rydberg interactions},
	url = {https://doi.org/10.1038/s41467-023-42899-8},
	volume = {14},
	year = {2023}
}

@article{Koyluoglu2025,
	author = {Koyluoglu, Nazli Ugur and Maskara, Nishad and Feldmeier, Johannes and Lukin, Mikhail D.},
	doi = {10.1103/5qhh-322q},
	issue = {11},
	journal = {Phys. Rev. Lett.},
	month = {Sep},
	numpages = {9},
	pages = {113603},
	publisher = {American Physical Society},
	title = {Floquet Engineering of Interactions and Entanglement in Periodically Driven Rydberg Chains},
	url = {https://link.aps.org/doi/10.1103/5qhh-322q},
	volume = {135},
	year = {2025}}

@article{Kalinowski2023,
	author = {Kalinowski, Marcin and Maskara, Nishad and Lukin, Mikhail D.},
	doi = {10.1103/PhysRevX.13.031008},
	issue = {3},
	journal = {Phys. Rev. X},
	month = {Jul},
	numpages = {22},
	pages = {031008},
	publisher = {American Physical Society},
	title = {Non-Abelian Floquet Spin Liquids in a Digital Rydberg Simulator},
	url = {https://link.aps.org/doi/10.1103/PhysRevX.13.031008},
	volume = {13},
	year = {2023}
}

@article{Greif2013,
	author = {Daniel Greif and Thomas Uehlinger and Gregor Jotzu and Leticia Tarruell and Tilman Esslinger},
	doi = {10.1126/science.1236362},
	journal = {Science},
	number = {6138},
	pages = {1307-1310},
	title = {Short-Range Quantum Magnetism of Ultracold Fermions in an Optical Lattice},
	url = {https://www.science.org/doi/abs/10.1126/science.1236362},
	volume = {340},
	year = {2013}
}

@article{DiLiberto2014,
	author = {Liberto, M. Di and Creffield, C. E. and Japaridze, G. I. and Smith, C. Morais},
	doi = {10.1103/PhysRevA.89.013624},
	issue = {1},
	journal = {Phys. Rev. A},
	month = {Jan},
	numpages = {10},
	pages = {013624},
	publisher = {American Physical Society},
	title = {Quantum simulation of correlated-hopping models with fermions in optical lattices},
	url = {https://link.aps.org/doi/10.1103/PhysRevA.89.013624},
	volume = {89},
	year = {2014}
}

@article{Rapp2012,
	author = {Rapp, \'Akos and Deng, Xiaolong and Santos, Luis},
	doi = {10.1103/PhysRevLett.109.203005},
	issue = {20},
	journal = {Phys. Rev. Lett.},
	month = {Nov},
	numpages = {5},
	pages = {203005},
	publisher = {American Physical Society},
	title = {Ultracold Lattice Gases with Periodically Modulated Interactions},
	url = {https://link.aps.org/doi/10.1103/PhysRevLett.109.203005},
	volume = {109},
	year = {2012}
}

@article{Aidelsburger2011,
	author = {Aidelsburger, M. and Atala, M. and Nascimb\`ene, S. and Trotzky, S. and Chen, Y.-A. and Bloch, I.},
	doi = {10.1103/PhysRevLett.107.255301},
	issue = {25},
	journal = {Phys. Rev. Lett.},
	month = {Dec},
	numpages = {5},
	pages = {255301},
	publisher = {American Physical Society},
	title = {Experimental Realization of Strong Effective Magnetic Fields in an Optical Lattice},
	url = {https://link.aps.org/doi/10.1103/PhysRevLett.107.255301},
	volume = {107},
	year = {2011}
}

@article{Goldman2014,
	author = {Goldman, N and Juzeli{\=u}nas, G and {\"O}hberg, P and Spielman, I B},
	doi = {10.1088/0034-4885/77/12/126401},
	journal = {Reports on Progress in Physics},
	month = {nov},
	number = {12},
	pages = {126401},
	publisher = {IOP Publishing},
	title = {Light-induced gauge fields for ultracold atoms},
	url = {https://doi.org/10.1088/0034-4885/77/12/126401},
	volume = {77},
	year = {2014}
}

@article{Goldman2016,
	author = {Goldman, N. and Budich, J. C. and Zoller, P.},
	date = {2016/07/01},
	doi = {10.1038/nphys3803},
	id = {Goldman2016},
	isbn = {1745-2481},
	journal = {Nature Physics},
	number = {7},
	pages = {639--645},
	title = {Topological quantum matter with ultracold gases in optical lattices},
	url = {https://doi.org/10.1038/nphys3803},
	volume = {12},
	year = {2016}
}

@article{Zenesini2009,
	author = {Zenesini, Alessandro and Lignier, Hans and Ciampini, Donatella and Morsch, Oliver and Arimondo, Ennio},
	doi = {10.1103/PhysRevLett.102.100403},
	issue = {10},
	journal = {Phys. Rev. Lett.},
	month = {Mar},
	numpages = {4},
	pages = {100403},
	publisher = {American Physical Society},
	title = {Coherent Control of Dressed Matter Waves},
	url = {https://link.aps.org/doi/10.1103/PhysRevLett.102.100403},
	volume = {102},
	year = {2009}}

@article{Eckardt2005,
	author = {Eckardt, Andr\'e and Weiss, Christoph and Holthaus, Martin},
	doi = {10.1103/PhysRevLett.95.260404},
	issue = {26},
	journal = {Phys. Rev. Lett.},
	month = {Dec},
	numpages = {4},
	pages = {260404},
	publisher = {American Physical Society},
	title = {Superfluid-Insulator Transition in a Periodically Driven Optical Lattice},
	url = {https://link.aps.org/doi/10.1103/PhysRevLett.95.260404},
	volume = {95},
	year = {2005}}

@article{Cooper2019,
	author = {Cooper, N. R. and Dalibard, J. and Spielman, I. B.},
	doi = {10.1103/RevModPhys.91.015005},
	issue = {1},
	journal = {Rev. Mod. Phys.},
	month = {Mar},
	numpages = {55},
	pages = {015005},
	publisher = {American Physical Society},
	title = {Topological bands for ultracold atoms},
	url = {https://link.aps.org/doi/10.1103/RevModPhys.91.015005},
	volume = {91},
	year = {2019}
}

@article{Eckardt2017,
	author = {Eckardt, Andr\'e},
	doi = {10.1103/RevModPhys.89.011004},
	issue = {1},
	journal = {Rev. Mod. Phys.},
	month = {Mar},
	numpages = {30},
	pages = {011004},
	publisher = {American Physical Society},
	title = {Colloquium: Atomic quantum gases in periodically driven optical lattices},
	url = {https://link.aps.org/doi/10.1103/RevModPhys.89.011004},
	volume = {89},
	year = {2017}
}

@article{Santos2021,
	author = {Li, Wei-Han and Dhar, Arya and Deng, Xiaolong and Santos, Luis},
	doi = {10.1103/PhysRevA.103.043331},
	issue = {4},
	journal = {Phys. Rev. A},
	month = {Apr},
	numpages = {8},
	pages = {043331},
	publisher = {American Physical Society},
	title = {Cluster dynamics in two-dimensional lattice gases with intersite interactions},
	url = {https://link.aps.org/doi/10.1103/PhysRevA.103.043331},
	volume = {103},
	year = {2021}
}

@article{Salerno2020,
	author = {Salerno, G. and Palumbo, G. and Goldman, N. and Di Liberto, M.},
	doi = {10.1103/PhysRevResearch.2.013348},
	issue = {1},
	journal = {Phys. Rev. Res.},
	month = {Mar},
	numpages = {9},
	pages = {013348},
	publisher = {American Physical Society},
	title = {Interaction-induced lattices for bound states: Designing flat bands, quantized pumps, and higher-order topological insulators for doublons},
	url = {https://link.aps.org/doi/10.1103/PhysRevResearch.2.013348},
	volume = {2},
	year = {2020}}

@article{Valiente2009,
	author = {Valiente, M and Petrosyan, D},
	doi = {10.1088/0953-4075/42/12/121001},
	journal = {Journal of Physics B: Atomic, Molecular and Optical Physics},
	month = {jun},
	number = {12},
	pages = {121001},
	title = {Scattering resonances and two-particle bound states of the extended Hubbard model},
	url = {https://doi.org/10.1088/0953-4075/42/12/121001},
	volume = {42},
	year = {2009}
}

@article{Winkler2006,
	author = {Winkler, K. and Thalhammer, G. and Lang, F. and Grimm, R. and Hecker Denschlag, J. and Daley, A. J. and Kantian, A. and B{\"u}chler, H. P. and Zoller, P.},
	date = {2006/06/01},
	doi = {10.1038/nature04918},
	id = {Winkler2006},
	isbn = {1476-4687},
	journal = {Nature},
	number = {7095},
	pages = {853--856},
	title = {Repulsively bound atom pairs in an optical lattice},
	url = {https://doi.org/10.1038/nature04918},
	volume = {441},
	year = {2006}
}

@article{Lukin2021,
	author = {Ebadi, Sepehr and Wang, Tout T. and Levine, Harry and Keesling, Alexander and Semeghini, Giulia and Omran, Ahmed and Bluvstein, Dolev and Samajdar, Rhine and Pichler, Hannes and Ho, Wen Wei and Choi, Soonwon and Sachdev, Subir and Greiner, Markus and Vuleti{\'c}, Vladan and Lukin, Mikhail D.},
	date = {2021/07/01},
	doi = {10.1038/s41586-021-03582-4},
	id = {Ebadi2021},
	isbn = {1476-4687},
	journal = {Nature},
	number = {7866},
	pages = {227--232},
	title = {Quantum phases of matter on a 256-atom programmable quantum simulator},
	url = {https://doi.org/10.1038/s41586-021-03582-4},
	volume = {595},
	year = {2021}
}

@book{sachdev2023,
  title={Quantum phases of matter},
  author={Sachdev, Subir},
  year={2023},
  publisher={Cambridge University Press},
  url = {https://www.cambridge.org/core/books/quantum-phases-of-matter/1D3F53A6FB1B448CE2484C5F797A1A00}
}

@article{Georgescu2014,
	author = {Georgescu, I. M. and Ashhab, S. and Nori, Franco},
	doi = {10.1103/RevModPhys.86.153},
	issue = {1},
	journal = {Rev. Mod. Phys.},
	month = {Mar},
	numpages = {33},
	pages = {153--185},
	publisher = {American Physical Society},
	title = {Quantum simulation},
	url = {https://link.aps.org/doi/10.1103/RevModPhys.86.153},
	volume = {86},
	year = {2014}
}

@article{Baranov2012,
	author = {Baranov, M. A. and Dalmonte, M. and Pupillo, G. and Zoller, P.},
	date = {2012/09/12},
	doi = {10.1021/cr2003568},
	isbn = {0009-2665},
	journal = {Chemical Reviews},
	journal1 = {Chemical Reviews},
	journal2 = {Chem. Rev.},
	month = {09},
	number = {9},
	pages = {5012--5061},
	publisher = {American Chemical Society},
	title = {Condensed Matter Theory of Dipolar Quantum Gases},
	type = {doi: 10.1021/cr2003568},
	url = {https://doi.org/10.1021/cr2003568},
	volume = {112},
	year = {2012},
	year1 = {2012}
}

@article{Trefzger2011,
	author = {Trefzger, C and Menotti, C and Capogrosso-Sansone, B and Lewenstein, M},
	doi = {10.1088/0953-4075/44/19/193001},
	journal = {Journal of Physics B: Atomic, Molecular and Optical Physics},
	month = {sep},
	number = {19},
	pages = {193001},
	title = {Ultracold dipolar gases in optical lattices},
	url = {https://doi.org/10.1088/0953-4075/44/19/193001},
	volume = {44},
	year = {2011}
}

@article{Greiner2017,
	author = {Mazurenko, Anton and Chiu, Christie S. and Ji, Geoffrey and Parsons, Maxwell F. and Kan{\'a}sz-Nagy, M{\'a}rton and Schmidt, Richard and Grusdt, Fabian and Demler, Eugene and Greif, Daniel and Greiner, Markus},
	date = {2017/05/01},
	doi = {10.1038/nature22362},
	id = {Mazurenko2017},
	isbn = {1476-4687},
	journal = {Nature},
	number = {7655},
	pages = {462--466},
	title = {A cold-atom Fermi--Hubbard antiferromagnet},
	url = {https://doi.org/10.1038/nature22362},
	volume = {545},
	year = {2017}
}

@article{Gross2016,
	author = {Martin Boll and Timon A. Hilker and Guillaume Salomon and Ahmed Omran and Jacopo Nespolo and Lode Pollet and Immanuel Bloch and Christian Gross},
	doi = {10.1126/science.aag1635},
	journal = {Science},
	number = {6305},
	pages = {1257-1260},
	title = {Spin- and density-resolved microscopy of antiferromagnetic correlations in Fermi-Hubbard chains},
	url = {https://www.science.org/doi/abs/10.1126/science.aag1635},
	volume = {353},
	year = {2016}
}

@article{Gross2013,
	author = {Fukuhara, Takeshi and Schau{\ss}, Peter and Endres, Manuel and Hild, Sebastian and Cheneau, Marc and Bloch, Immanuel and Gross, Christian},
	date = {2013/10/01},
	doi = {10.1038/nature12541},
	id = {Fukuhara2013},
	isbn = {1476-4687},
	journal = {Nature},
	number = {7469},
	pages = {76--79},
	title = {Microscopic observation of magnon bound states and their dynamics},
	url = {https://doi.org/10.1038/nature12541},
	volume = {502},
	year = {2013}
}

@article{Duan2003,
  title = {Controlling Spin Exchange Interactions of Ultracold Atoms in Optical Lattices},
  author = {Duan, L.-M. and Demler, E. and Lukin, M. D.},
  journal = {Phys. Rev. Lett.},
  volume = {91},
  issue = {9},
  pages = {090402},
  numpages = {4},
  year = {2003},
  month = {Aug},
  publisher = {American Physical Society},
  doi = {10.1103/PhysRevLett.91.090402},
  url = {https://link.aps.org/doi/10.1103/PhysRevLett.91.090402}
}

@article{de_leseleuc_observation_2019,
	title = {Observation of a symmetry-protected topological phase of interacting bosons with {Rydberg} atoms},
	volume = {365},
	issn = {0036-8075, 1095-9203},
	url = {https://www.science.org/doi/10.1126/science.aav9105},
	doi = {10.1126/science.aav9105},
	language = {en},
	number = {6455},
	urldate = {2023-04-11},
	journal = {Science},
	author = {de Léséleuc, Sylvain and Lienhard, Vincent and Scholl, Pascal and Barredo, Daniel and Weber, Sebastian and Lang, Nicolai and Büchler, Hans Peter and Lahaye, Thierry and Browaeys, Antoine},
	month = aug,
	year = {2019},
	pages = {775--780},
}

@article{scholl_microwave-engineering_2022,
  title={Microwave engineering of programmable XXZ Hamiltonians in arrays of Rydberg atoms},
  author={Scholl, Pascal and Williams, Hannah J and Bornet, Guillaume and Wallner, Florian and Barredo, Daniel and Henriet, L and Signoles, Adrien and Hainaut, Cl{\'e}ment and Franz, Titus and Geier, Sebastian and others},
  journal={PRX Quantum},
  volume={3},
  number={2},
  pages={020303},
  year={2022},
  publisher={APS},
  url = {https://journals.aps.org/prxquantum/abstract/10.1103/PRXQuantum.3.020303}
}

@article{geier_floquet_2021,
	title = {Floquet {Hamiltonian} engineering of an isolated many-body spin system},
	volume = {374},
	issn = {0036-8075, 1095-9203},
	url = {https://www.science.org/doi/10.1126/science.abd9547},
	doi = {10.1126/science.abd9547},
	language = {en},
	number = {6571},
	urldate = {2023-10-27},
	journal = {Science},
	author = {Geier, Sebastian and Thaicharoen, Nithiwadee and Hainaut, Clément and Franz, Titus and Salzinger, Andre and Tebben, Annika and Grimshandl, David and Zürn, Gerhard and Weidemüller, Matthias},
	month = nov,
	year = {2021},
	pages = {1149--1152},
}

@article{goldman_periodically_2014,
	title = {Periodically {Driven} {Quantum} {Systems}: {Effective} {Hamiltonians} and {Engineered} {Gauge} {Fields}},
	volume = {4},
	issn = {2160-3308},
	shorttitle = {Periodically {Driven} {Quantum} {Systems}},
	url = {https://link.aps.org/doi/10.1103/PhysRevX.4.031027},
	doi = {10.1103/PhysRevX.4.031027},
	language = {en},
	number = {3},
	urldate = {2023-11-08},
	journal = {Physical Review X},
	author = {Goldman, N. and Dalibard, J.},
	month = aug,
	year = {2014},
	pages = {031027},
}

@article{chen_spectroscopy_2023,
author = {Cheng Chen  and Gabriel Emperauger  and Guillaume Bornet  and Filippo Caleca  and Bastien Gély  and Marcus Bintz  and Shubhayu Chatterjee  and Vincent Liu  and Daniel Barredo  and Norman Y. Yao  and Thierry Lahaye  and Fabio Mezzacapo  and Tommaso Roscilde  and Antoine Browaeys },
title = {Spectroscopy of elementary excitations from quench dynamics in a dipolar XY Rydberg simulator},
journal = {Science},
volume = {389},
number = {6759},
pages = {483-487},
year = {2025},
doi = {10.1126/science.adn0618},
URL = {https://www.science.org/doi/abs/10.1126/science.adn0618}}

@article{lienhard_realization_2020,
	title = {Realization of a {Density}-{Dependent} {Peierls} {Phase} in a {Synthetic}, {Spin}-{Orbit} {Coupled} {Rydberg} {System}},
	volume = {10},
	issn = {2160-3308},
	url = {https://link.aps.org/doi/10.1103/PhysRevX.10.021031},
	doi = {10.1103/PhysRevX.10.021031},
	language = {en},
	number = {2},
	urldate = {2024-07-26},
	journal = {Physical Review X},
	author = {Lienhard, Vincent and Scholl, Pascal and Weber, Sebastian and Barredo, Daniel and De Léséleuc, Sylvain and Bai, Rukmani and Lang, Nicolai and Fleischhauer, Michael and Büchler, Hans Peter and Lahaye, Thierry and Browaeys, Antoine},
	month = may,
	year = {2020},
	pages = {021031},
}

@article{ravets_measurement_2015,
  title = {Measurement of the angular dependence of the dipole-dipole interaction between two individual Rydberg atoms at a F\"orster resonance},
  author = {Ravets, Sylvain and Labuhn, Henning and Barredo, Daniel and Lahaye, Thierry and Browaeys, Antoine},
  journal = {Phys. Rev. A},
  volume = {92},
  issue = {2},
  pages = {020701},
  numpages = {5},
  year = {2015},
  month = {Aug},
  publisher = {American Physical Society},
  doi = {10.1103/PhysRevA.92.020701},
  url = {https://link.aps.org/doi/10.1103/PhysRevA.92.020701}
}

@article{saffman_quantum_2010,
	title = {Quantum information with {Rydberg} atoms},
	volume = {82},
	copyright = {http://link.aps.org/licenses/aps-default-license},
	issn = {0034-6861, 1539-0756},
	url = {https://link.aps.org/doi/10.1103/RevModPhys.82.2313},
	doi = {10.1103/RevModPhys.82.2313},
	language = {en},
	number = {3},
	urldate = {2024-07-28},
	journal = {Reviews of Modern Physics},
	author = {Saffman, M. and Walker, T. G. and Mølmer, K.},
	month = aug,
	year = {2010},
	pages = {2313--2363},
}

@article{Semeghini2021,
	author = {G. Semeghini and H. Levine and A. Keesling and S. Ebadi and T. T. Wang and D. Bluvstein and R. Verresen and H. Pichler and M. Kalinowski and R. Samajdar and A. Omran and S. Sachdev and A. Vishwanath and M. Greiner and V. Vuleti{\'c} and M. D. Lukin},
	doi = {10.1126/science.abi8794},
	journal = {Science},
	number = {6572},
	pages = {1242-1247},
	title = {Probing topological spin liquids on a programmable quantum simulator},
	url = {https://www.science.org/doi/abs/10.1126/science.abi8794},
	volume = {374},
	year = {2021}
}

@article{weckesser_realization_2024,
  title={Realization of a Rydberg-dressed extended Bose-Hubbard model},
  author={Weckesser, Pascal and Srakaew, Kritsana and Blatz, Tizian and Wei, David and Adler, Daniel and Agrawal, Suchita and Bohrdt, Annabelle and Bloch, Immanuel and Zeiher, Johannes},
  journal={Science},
  volume={390},
  number={6775},
  pages={849--853},
  year={2025},
  publisher={American Association for the Advancement of Science},
  url = {https://www.science.org/doi/10.1126/science.adq7082}
}

@article{de_leseleuc_optical_2017,
	title = {Optical {Control} of the {Resonant} {Dipole}-{Dipole} {Interaction} between {Rydberg} {Atoms}},
	volume = {119},
	copyright = {http://link.aps.org/licenses/aps-default-license},
	issn = {0031-9007, 1079-7114},
	url = {http://link.aps.org/doi/10.1103/PhysRevLett.119.053202},
	doi = {10.1103/PhysRevLett.119.053202},
	language = {en},
	number = {5},
	urldate = {2024-11-20},
	journal = {Physical Review Letters},
	author = {De Léséleuc, Sylvain and Barredo, Daniel and Lienhard, Vincent and Browaeys, Antoine and Lahaye, Thierry},
	month = aug,
	year = {2017},
	pages = {053202},
}

@article{tai_microscopy_2017,
	title = {Microscopy of the interacting {Harper}–{Hofstadter} model in the two-body limit},
	volume = {546},
	issn = {0028-0836, 1476-4687},
	url = {https://www.nature.com/articles/nature22811},
	doi = {10.1038/nature22811},
	language = {en},
	number = {7659},
	urldate = {2024-12-12},
	journal = {Nature},
	author = {Tai, M. Eric and Lukin, Alexander and Rispoli, Matthew and Schittko, Robert and Menke, Tim and {Dan Borgnia} and Preiss, Philipp M. and Grusdt, Fabian and Kaufman, Adam M. and Greiner, Markus},
	month = jun,
	year = {2017},
	pages = {519--523},
}

@article{Evered2025,
  title={Probing the Kitaev honeycomb model on a neutral-atom quantum computer},
  author={Evered, Simon J and Kalinowski, Marcin and Geim, Alexandra A and Manovitz, Tom and Bluvstein, Dolev and Li, Sophie H and Maskara, Nishad and Zhou, Hengyun and Ebadi, Sepehr and Xu, Muqing and others},
  journal={Nature},
  volume={645},
  number={8080},
  pages={341--347},
  year={2025},
  publisher={Nature Publishing Group UK London},
  url = {https://www.nature.com/articles/s41586-025-09475-0}
}

@article{rapp_ultracold_2012,
	title = {Ultracold {Lattice} {Gases} with {Periodically} {Modulated} {Interactions}},
	volume = {109},
	copyright = {http://link.aps.org/licenses/aps-default-license},
	issn = {0031-9007, 1079-7114},
	url = {https://link.aps.org/doi/10.1103/PhysRevLett.109.203005},
	doi = {10.1103/PhysRevLett.109.203005},
	language = {en},
	number = {20},
	urldate = {2025-03-05},
	journal = {Physical Review Letters},
	author = {Rapp, \'Akos and Deng, Xiaolong and Santos, Luis},
	month = nov,
	year = {2012},
	pages = {203005},
}

@article{preiss_strongly_nodate,
  title={Strongly correlated quantum walks in optical lattices},
  author={Preiss, Philipp M and Ma, Ruichao and Tai, M Eric and Lukin, Alexander and Rispoli, Matthew and Zupancic, Philip and Lahini, Yoav and Islam, Rajibul and Greiner, Markus},
  journal={Science},
  volume={347},
  number={6227},
  pages={1229--1233},
  year={2015},
  publisher={American Association for the Advancement of Science},
  url = {https://www.science.org/doi/10.1126/science.1260364}
}

@article{browaeys_many-body_2020,
	title = {Many-body physics with individually controlled {Rydberg} atoms},
	volume = {16},
	copyright = {2020 Springer Nature Limited},
	issn = {1745-2481},
	url = {https://www.nature.com/articles/s41567-019-0733-z},
	doi = {10.1038/s41567-019-0733-z},
	language = {en},
	number = {2},
	urldate = {2025-06-07},
	journal = {Nature Physics},
	author = {Browaeys, Antoine and Lahaye, Thierry},
	month = feb,
	year = {2020},
	note = {Publisher: Nature Publishing Group},
	pages = {132--142},
}

@article{scholl_quantum_2021,
	title = {Quantum simulation of {2D} antiferromagnets with hundreds of {Rydberg} atoms},
	volume = {595},
	copyright = {2021 The Author(s), under exclusive licence to Springer Nature Limited},
	issn = {1476-4687},
	url = {https://www.nature.com/articles/s41586-021-03585-1},
	doi = {10.1038/s41586-021-03585-1},
	language = {en},
	number = {7866},
	urldate = {2025-06-07},
	journal = {Nature},
	author = {Scholl, Pascal and Schuler, Michael and Williams, Hannah J. and Eberharter, Alexander A. and Barredo, Daniel and Schymik, Kai-Niklas and Lienhard, Vincent and Henry, Louis-Paul and Lang, Thomas C. and Lahaye, Thierry and Läuchli, Andreas M. and Browaeys, Antoine},
	month = jul,
	year = {2021},
	note = {Publisher: Nature Publishing Group},
	pages = {233--238},
}

@article{waugh_approach_1968,
	title = {Approach to {High}-{Resolution} nmr in {Solids}},
	volume = {20},
	copyright = {http://link.aps.org/licenses/aps-default-license},
	issn = {0031-9007},
	url = {https://link.aps.org/doi/10.1103/PhysRevLett.20.180},
	doi = {10.1103/PhysRevLett.20.180},
	language = {en},
	number = {5},
	urldate = {2025-06-12},
	journal = {Physical Review Letters},
	author = {Waugh, J. S. and Huber, L. M. and Haeberlen, U.},
	month = jan,
	year = {1968},
	pages = {180--182},
}

@article{macri_bound_2021,
	title = {Bound state dynamics in the long-range spin- 1 2 {XXZ} model},
	volume = {104},
	issn = {2469-9950, 2469-9969},
	url = {https://link.aps.org/doi/10.1103/PhysRevB.104.214309},
	doi = {10.1103/PhysRevB.104.214309},
	language = {en},
	number = {21},
	urldate = {2025-09-02},
	journal = {Physical Review B},
	author = {Macrì, T. and Lepori, L. and Pagano, G. and Lewenstein, M. and Barbiero, L.},
	month = dec,
	year = {2021},
	pages = {214309},
}

@article{Manovitz2025,
	author = {Manovitz, Tom and Li, Sophie H. and Ebadi, Sepehr and Samajdar, Rhine and Geim, Alexandra A. and Evered, Simon J. and Bluvstein, Dolev and Zhou, Hengyun and Koyluoglu, Nazli Ugur and Feldmeier, Johannes and Dolgirev, Pavel E. and Maskara, Nishad and Kalinowski, Marcin and Sachdev, Subir and Huse, David A. and Greiner, Markus and Vuleti{\ifmmode\acute{c}\else\'{c}\fi}, Vladan and Lukin, Mikhail D.},
	title = {{Quantum coarsening and collective dynamics on a programmable simulator}},
	journal = {Nature},
	volume = {638},
	pages = {86--92},
	year = {2025},
	month = feb,
	issn = {1476-4687},
	publisher = {Nature Publishing Group},
	doi = {10.1038/s41586-024-08353-5}
}

@article{Bernien2017,
	author = {Bernien, Hannes and Schwartz, Sylvain and Keesling, Alexander and Levine, Harry and Omran, Ahmed and Pichler, Hannes and Choi, Soonwon and Zibrov, Alexander S. and Endres, Manuel and Greiner, Markus and Vuleti{\ifmmode\acute{c}\else\'{c}\fi}, Vladan and Lukin, Mikhail D.},
	title = {{Probing many-body dynamics on a 51-atom quantum simulator}},
	journal = {Nature},
	volume = {551},
	pages = {579--584},
	year = {2017},
	month = nov,
	issn = {1476-4687},
	publisher = {Nature Publishing Group},
	doi = {10.1038/nature24622}
}

@article{abanin_2015,
  title = {Exponentially Slow Heating in Periodically Driven Many-Body Systems},
  author = {Abanin, Dmitry A. and De Roeck, Wojciech and Huveneers, Fran\ifmmode \mbox{\c{c}}\else \c{c}\fi{}ois},
  journal = {Phys. Rev. Lett.},
  volume = {115},
  issue = {25},
  pages = {256803},
  numpages = {5},
  year = {2015},
  month = {Dec},
  publisher = {American Physical Society},
  doi = {10.1103/PhysRevLett.115.256803},
  url = {https://link.aps.org/doi/10.1103/PhysRevLett.115.256803}
}

@article{ganahl_2012,
  title = {Observation of Complex Bound States in the Spin-$1/2$ Heisenberg $XXZ$ Chain Using Local Quantum Quenches},
  author = {Ganahl, Martin and Rabel, Elias and Essler, Fabian H. L. and Evertz, H. G.},
  journal = {Phys. Rev. Lett.},
  volume = {108},
  issue = {7},
  pages = {077206},
  numpages = {5},
  year = {2012},
  month = {Feb},
  publisher = {American Physical Society},
  doi = {10.1103/PhysRevLett.108.077206},
  url = {https://link.aps.org/doi/10.1103/PhysRevLett.108.077206}
}

@article{FossFeig2015,
  title = {Nearly Linear Light Cones in Long-Range Interacting Quantum Systems},
  author = {Foss-Feig, Michael and Gong, Zhe-Xuan and Clark, Charles W. and Gorshkov, Alexey V.},
  journal = {Phys. Rev. Lett.},
  volume = {114},
  issue = {15},
  pages = {157201},
  numpages = {5},
  year = {2015},
  month = {Apr},
  publisher = {American Physical Society},
  doi = {10.1103/PhysRevLett.114.157201},
  url = {https://link.aps.org/doi/10.1103/PhysRevLett.114.157201}
}

@article{valiente_2010,
  title = {Three-body bound states in a lattice},
  author = {Valiente, Manuel and Petrosyan, David and Saenz, Alejandro},
  journal = {Phys. Rev. A},
  volume = {81},
  issue = {1},
  pages = {011601},
  numpages = {4},
  year = {2010},
  month = {Jan},
  publisher = {American Physical Society},
  doi = {10.1103/PhysRevA.81.011601},
  url = {https://link.aps.org/doi/10.1103/PhysRevA.81.011601}
}

@article{kim_2024,
  title = {Realization of an Extremely Anisotropic Heisenberg Magnet in Rydberg Atom Arrays},
  author = {Kim, Kangheun and Yang, Fan and M\o{}lmer, Klaus and Ahn, Jaewook},
  journal = {Phys. Rev. X},
  volume = {14},
  issue = {1},
  pages = {011025},
  numpages = {13},
  year = {2024},
  month = {Feb},
  publisher = {American Physical Society},
  doi = {10.1103/PhysRevX.14.011025},
  url = {https://link.aps.org/doi/10.1103/PhysRevX.14.011025}
}

@article{giudice_2022,
  title = {Trimer states with ${\mathbb{Z}}_{3}$ topological order in Rydberg atom arrays},
  author = {Giudice, Giacomo and Surace, Federica Maria and Pichler, Hannes and Giudici, Giuliano},
  journal = {Phys. Rev. B},
  volume = {106},
  issue = {19},
  pages = {195155},
  numpages = {15},
  year = {2022},
  month = {Nov},
  publisher = {American Physical Society},
  doi = {10.1103/PhysRevB.106.195155},
  url = {https://link.aps.org/doi/10.1103/PhysRevB.106.195155}
}

@article{kranzl_2023,
  title = {Observation of Magnon Bound States in the Long-Range, Anisotropic Heisenberg Model},
  author = {Kranzl, Florian and Birnkammer, Stefan and Joshi, Manoj K. and Bastianello, Alvise and Blatt, Rainer and Knap, Michael and Roos, Christian F.},
  journal = {Phys. Rev. X},
  volume = {13},
  issue = {3},
  pages = {031017},
  numpages = {12},
  year = {2023},
  month = {Aug},
  publisher = {American Physical Society},
  doi = {10.1103/PhysRevX.13.031017},
  url = {https://link.aps.org/doi/10.1103/PhysRevX.13.031017}
}

@article{misguich_2017,
  title = {Dynamics of the spin-$\frac{1}{2}$ Heisenberg chain initialized in a domain-wall state},
  author = {Misguich, Gr\'egoire and Mallick, Kirone and Krapivsky, P. L.},
  journal = {Phys. Rev. B},
  volume = {96},
  issue = {19},
  pages = {195151},
  numpages = {7},
  year = {2017},
  month = {Nov},
  publisher = {American Physical Society},
  doi = {10.1103/PhysRevB.96.195151},
  url = {https://link.aps.org/doi/10.1103/PhysRevB.96.195151}
}

@article{weinberg2017quspin,
  title={QuSpin: a Python package for dynamics and exact diagonalisation of quantum many body systems part I: spin chains},
  author={Weinberg, Phillip and Bukov, Marin},
  journal={SciPost Physics},
  volume={2},
  number={1},
  pages={003},
  year={2017}
}

@article{weinberg2019quspin,
  title={QuSpin: a Python package for dynamics and exact diagonalisation of quantum many body systems part II: bosons, fermions and higher spins},
  author={Weinberg, Phillip and Bukov, Marin},
  journal={SciPost Physics},
  volume={7},
  number={2},
  pages={020},
  year={2019}
}

\end{document}